\documentclass[]{aastex631}

\newcommand{\kms}{\rm km~s^{-1}}
\newcommand{\zcat}{z^{\rm SDSScat}}
\newcommand{\zpysdsssdss}{z^{\rm RVSNUpy}_{\rm SDSS, SDSS}}
\newcommand{\zpyhectosdss}{z^{\rm RVSNUpy}_{\rm Hecto, SDSS}}

\newcommand{\zpyhectohecto}{z^{\rm RVSNUpy}_{\rm Hecto, Hecto}}
\newcommand{\zpysdssutemp}{z^{\rm RVSNUpy}_{\rm SDSS, utemp}}
\newcommand{\zpyhectoutemp}{z^{\rm RVSNUpy}_{\rm Hecto, utemp}}
\newcommand{\zpysdssfsps}{z^{\rm RVSNUpy}_{\rm SDSS, FSPS}}

\usepackage{color}
\usepackage{amsmath}
\usepackage{multirow}

\begin{document}

\title{RVSNUpy: A Python Package for Spectroscopic Redshift Measurement Based on Cross-Correlation}

\correspondingauthor{Jubee Sohn}
\email{jubee.sohn@snu.ac.kr} 

\author[0009-0001-3758-9440]{Taewan Kim}
\affiliation{Astronomy Program, Department of Physics and Astronomy, Seoul National University, 1 Gwanak-ro, Gwanak-gu, Seoul 08826, Republic of Korea} 
\affiliation{SNU Astronomy Research Center, Seoul National University, Seoul 08826, Republic of Korea}

\author[0000-0002-9254-144X]{Jubee Sohn}
\affiliation{Astronomy Program, Department of Physics and Astronomy, Seoul National University, 1 Gwanak-ro, Gwanak-gu, Seoul 08826, Republic of Korea} 
\affiliation{SNU Astronomy Research Center, Seoul National University, Seoul 08826, Republic of Korea} 

\author[0000-0003-3428-7612]{Ho Seong Hwang}
\affiliation{Astronomy Program, Department of Physics and Astronomy, Seoul National University, 1 Gwanak-ro, Gwanak-gu, Seoul 08826, Republic of Korea} 
\affiliation{SNU Astronomy Research Center, Seoul National University, Seoul 08826, Republic of Korea}
\affiliation{Australian Astronomical Optics - Macquarie University, 105 Delhi Road, North Ryde, NSW 2113, Australia} 

\begin{abstract}
We introduce \texttt{RVSNUpy}, a new Python package designed to measure spectroscopic redshifts. Based on inverse-variance weighted cross-correlation, \texttt{RVSNUpy} determines the redshifts by comparing observed spectra with various rest-frame template spectra. We test the performance of \texttt{RVSNUpy} based on $\sim 6000$ objects in the HectoMAP redshift survey observed with both SDSS and MMT/Hectospec. We demonstrate that a slight redshift offset ($\sim 40~\kms$) between SDSS and MMT/Hectospec measurements reported from previous studies results from the small offsets in the redshift template spectra used for SDSS and Hectospec reductions. We construct the universal set of template spectra, including empirical SDSS template spectra, carefully calibrated to the rest frame. Our test for the HectoMAP objects with duplicated observations shows that \texttt{RVSNUpy} with the universal template spectra yields the homogeneous redshift from the spectra obtained with different spectrographs. We highlight that \texttt{RVSNUpy} is a powerful redshift measurement tool for current and future large-scale spectroscopy surveys, including A-SPEC, DESI, 4MOST, and Subaru/PFS. 
\end{abstract}


\section{Introduction} \label{sec:intro}

Redshift is a fundamental observable widely used in astrophysics and cosmology because it allows us to access three-dimensional information on objects. For nearby objects that are not significantly affected by the Hubble flow, redshift indicates the amount of Doppler shift due to the relative radial velocity between the observer and the objects. For extragalactic objects in the expanding universe, redshifts serve as a useful proxy for distance, reflecting the rate at which the universe is expanding.

Current and future large spectroscopic surveys (e.g., \citealp{Huchra83, Geller89, Colless01, sdss_dr4, sdss_dr18, Sohn21, Liske15, Baldry14, Greene22, desi24, euclid22, Schlieder24}) provide redshifts for a statistically large sample of stars, galaxies, and quasars. These spectroscopic surveys construct maps of the universe that enable studies on various subjects, including the kinematic structure of the Milky Way (e.g., \citealp{Bonaca17, Belokurov18, Mackereth19, Belokurov20}), the dynamical structure and evolution of galaxies and their systems (e.g., \citealp{Baldry12, Blake13, Maraston13}), and large-scale structures as cosmological probes (e.g., \citealp{Eisenstein05, Beutler11, desi24b}).

Spectroscopic surveys use various methodologies to measure redshifts. Early surveys measured redshifts by identifying prominent spectral lines (e.g., \citealp{Humason56, Sandage78, Kirshner78}). Surveys began using more systematic measurement techniques as more spectra became available. The Center for Astrophysics (CfA) Redshift Surveys 1, 2 \citep{Huchra83, Geller89} measure redshifts by performing cross-correlation with rest-frame template spectra. The 2-degree-Field Galaxy Redshift (2dF) survey \citep{Colless01} uses direct line fitting and cross-correlation techniques in combination. Current and future large spectroscopic surveys use various methods for redshift measurements; the Sloan Digital Sky Survey (SDSS; \citealp{sdss_dr4, sdss_dr7, sdss_dr16, sdss_dr18}) exploits the cross-correlation technique in the early data releases and the full spectral fitting technique from Data Release 8 (DR8) that fits model spectra to observed spectra. The Galaxy And Mass Assembly (GAMA) \citep{Liske15, Baldry14, Baldry18} and HectoMAP surveys \citep{Geller15, Sohn21, Sohn23} adopt the cross-correlation technique. The Prime Focus Spectrograph (PFS) Galaxy Evolution Survey \citep{Greene22} and the Dark Energy Spectroscopic Instrument (DESI) surveys \citep{desi24} measure redshifts based on the full spectral fitting technique.

The systematic offset in redshift measurements is a critical issue when we combine redshifts from various surveys. Using different tools may result in inconsistencies in redshift measurements. For example, \citet{Sohn21} show that the redshift measurements of the HectoMAP survey \citep{Geller11, Hwang16, Sohn21, Sohn23} systematically deviate from those of SDSS. \citet{sdss_dr6} also show that SDSS redshift measurements between DR7 and DR8 display a systematic offset due to the SDSS pipeline updates that were performed between these two data releases.

To effectively combine redshifts from various surveys without suffering from systematic offset, a tool that measures redshifts homogeneously is necessary. We develop a new homogeneous redshift measurement Python package, \texttt{RVSNUpy} \citep{RVSNUpy}. Based on the cross-correlation technique, \texttt{RVSNUpy} measures redshifts by comparing spectral lines of observed spectra and rest frame template spectra. Because computing cross-correlation is mathematically straightforward and model-independent, the redshift measurement of \texttt{RVSNUpy} is free from systematic offsets introduced by complex fitting procedures.

\citet{Fellgett53, Fellgett55} were the first to suggest a technique to measure stellar radial velocity using cross-correlation, which was implemented by \citet{Griffin67}. \citet{Simkin74} formulated a method to measure redshifts by Fourier transforming the spectral fluxes and performing cross-correlation in Fourier space. \citet{Tonry79} constructed the theoretical groundwork for redshift measurements with the cross-correlation technique in Fourier space. Assuming that the spectral lines are perfect Gaussian lines, \citet{Tonry79} show that the cross-correlation signals have Gaussian shapes and derive redshifts and uncertainties from the cross-correlation signals. \texttt{RVSAO} \citep{Kurtz98} is a widely used IRAF package that measures redshifts based on the cross-correlation technique described in \citet{Tonry79}.

The redshift measurement package we develop here is based on the inverse variance weighted cross-correlation technique. Kelson et al. (2003)\footnote{https://code.obs.carnegiescience.edu/Algorithms} show that redshift measurements become more robust when we perform cross-correlation in real space than in Fourier space. When performing cross-correlation in Fourier space, we lose the information from flux uncertainties during the Fourier transformation. The loss of information from uncertainties indicates that noisy features in a spectrum can falsify cross-correlation signals with irrelevant spectral lines. In real space, we can include the impact of noise on the redshift measurement by weighting cross-correlation signals based on the relative ratios of flux uncertainties (i.e., inverse variances). Weighting spectra based on the flux uncertainty ratio prevents misidentifying noisy features as spectral lines. In addition, Kelson et al. (2003) show that incorporating the weighting factors simplifies the uncertainty estimation of the redshift measurement.

Here, we develop \texttt{RVSNUpy}, a Python package that derives spectroscopic redshifts based on the inverse variance weighted cross-correlation technique. In particular, we use the cross-correlation in real space following Kelson et al. (2003), in contrast to many other cross-correlation tools for measuring redshifts (e.g., \texttt{RVSAO}; \citealp{Kurtz98}). The cross-correlation in real space enables the reliable redshift measurement by incorporating the flux uncertainty and a mathematically robust redshift uncertainty estimation.

We describe the mathematical background of \texttt{RVSNUpy} in Section \ref{sec:background}. In Section \ref{sec:rvsnupy}, we explain the structure and algorithm of \texttt{RVSNUpy}. We describe the sets of templates we use in this paper in Section 4. We test the performance of \texttt{RVSNUpy} based on the various sets of synthetic spectra (Section \ref{sec:modeltest}). We also evaluate the performance of \texttt{RVSNUpy} based on various comparisons with previous redshift measurements based on the cross-correlation and the full spectral fitting methods in Section \ref{sec:obstest}. We conclude our development in Section \ref{sec:conclusion}.

\section{Mathematical Background} \label{sec:background}

Our new redshift measurement Python package, \texttt{RVSNUpy}, determines spectroscopic redshift by cross-correlating observed spectra with template spectra in real space. In this section, we review the mathematical background of cross-correlation. We describe redshift measurement based on cross-correlation in Section \ref{sec:cc}. In Section \ref{sec:uncertainty}, we introduce the method for estimating the uncertainties in the redshifts. In Section \ref{sec:significance}, we explore the significance of the cross-correlation signals.

\subsection{Redshift Measurement Based on Cross-correlation Technique} \label{sec:cc}

Cross-correlation is an operator that computes the similarity of two functions. A larger cross-correlation value indicates a higher degree of similarity between the functions. One of the many possible applications of cross-correlation is measurements of radial velocities of stars and redshifts of extragalactic objects based on the comparison between observed spectra and redshift template spectra. 

We apply a cross-correlation technique to measure redshifts from optical spectra. The cross-correlation technique requires a template spectrum in the rest-frame (i.e., $z = 0$). By shifting the template spectrum in the redshift space by $z'$, the cross-correlation process produces a signal that describes the relationship between the observed spectrum and template spectrum as a function of $z'$. The cross-correlation value reaches its maximum when the shift applied to the template spectrum matches the redshift of the observed spectrum. Thus, the redshift corresponding to the observed spectrum is determined by the value of $z'$ at the peak of the cross-correlation signal.

The cross-correlation between the observed and template spectra \citep{Tonry79} is:
\begin{equation} \label{eq:convencc}
C.C.(z') = \sum^{n}_{i=1}\left(\frac{G(\lambda_i)}{G'(\lambda_i)}-1\right)\left(\frac{T(z';\lambda_i)}{T'(z';\lambda_i)}-1\right) = \sum^{n}_{i=1}\left(\frac{G_i}{G'_i}-1\right)\left(\frac{T_i(z')}{T'_i(z'
)}-1\right). 
\end{equation}
Here, $G(\lambda_i)$ and $G'(\lambda_i)$ indicate the observed spectrum and its continuum (i.e., flux and continuum flux as a function of wavelength $\lambda_{i}$, respectively). $T(z';\lambda_i)$ and $T'(z';\lambda_i)$ are the template spectrum and its continuum. $T(z';\lambda_i)$ indicates the template spectrum shifted in wavelength by $z'$. For simplicity, we omit $\lambda_i$ from the notation and use the index $i$ to refer to the value at $\lambda_{i}$ (e.g., $G_{i}=G(\lambda_{i})$).

The shift of the template spectrum is straightforward when the spectrum is sampled in fixed intervals in log($\lambda$). The shift in the redshift space ($z'$) by wavelength shift ($\Delta \lambda$) is
\begin{equation} \label{eq:dlambda1}
z' = \frac{\Delta \lambda_{i}}{\lambda_{i}}.
\end{equation}
For spectra binned in constant logarithmic intervals, i.e., $\lambda_i=e^{C_{0} i +C_{1}}$ with constant $C_{0}$ and $C_{1}$, the shift of $\lambda_{i}$ to $\lambda_{i+\Delta i}$ is equivalent to the shift in the redshift space:
\begin{equation} \label{eq:dlambda2}
z'  = \frac{\lambda_{i+\Delta i}-\lambda_i}{\lambda_{i}} = e^{C_i\Delta i}.
\end{equation}
In this case, shifting the spectrum on a logarithmic scale results in a wavelength shift that directly corresponds to the redshift. In other words, if $G(\lambda_i)$, $G'_i$, $T_i(z')$ and $T'_i(z')$ in Equation \ref{eq:convencc} are all equally sampled in log scale, the cross-correlation signal we compute based on the wavelength shift is a function of redshift. The following argument in this section assumes $G_i$, $G'_i$, $T_i(z')$ and $T'_i(z')$ are equally-sampled on the log-scale. In practice, when using \texttt{RVSNUpy}, only $T_i(z')$ and $T'_i(z')$ need to be in log scale, because \texttt{RVSNUpy} shifts $T_i(z')$ along $z'$ and resamples $T_i(z')$ at every $z'$ to match the pixel scale of $G_i$ and $G'_i$ (see Section \ref{sec:process}).

To derive the redshift, we cross-correlate the observed spectrum with the template spectra. However, the observed spectrum and template spectrum often have different shapes, especially in terms of their continua. Galaxy spectra typically consist of a continuum and absorption/emission lines from various components, including stars, star-forming regions, the interstellar medium, and active galactic nuclei. Depending on the composition of a galaxy, the overall shape of the continuum in the observed spectrum may differ from that in the template spectra. These differences in continuum shapes can distort the cross-correlation signal. To mitigate this, we remove the continua from both the observed spectrum and the template spectra, allowing us to perform the cross-correlation without the influence of these continuum variations.

We normalize the observed and template spectra by their respective continua using Equation \ref{eq:convencc}. For example, $(G_{i}/G'_{i}) - 1$ indicates that we divide the observed spectrum ($G_{i}$) by its continuum ($G'_{i}$), and subtract one to remove the continuum, ensuring that the resulting continuum value is zero.
As a result, the cross-correlation signals described in Equation \ref{eq:convencc} are based solely on the spectral lines and exhibit a clear peak. In other words, normalizing the spectra ensures that we measure the redshift based on the shifts of the spectral lines. For example, the normalization of the spectra with S/N$>10$ generally reduces the width of a cross-correlation peak $7000-8000~\kms$ to $300-400 \kms$. We describe the practical approach to trace the continua of an observed spectrum and template spectra in Section \ref{sec:process}.

Here, we adopt an inverse-variance weighted cross-correlation, incorporating a weighting factor based on the signal-to-noise ratios (S/N) of the input spectrum. The weighted cross-correlation signal is described as follows (Kelson et al. 2003): 
\begin{equation} \label{eq:wcc} 
C.C.(z) = \sum^{n}_{i=1}\frac{M_i}{(\delta G_i/G_i')^2}\left(\frac{G_i}{G'_i}-1\right)\left(\frac{T_i(z')}{T'_i(z')}-1\right)=\sum^n_{i=1}\frac{M_i}{\delta G^2_i}(G_i-G_i')(T^\circ_i(z')-G'_i)\text{, where } T^\circ_i(z')=G'_i\frac{T_i(z')}{T'_i(z')}.
\end{equation}
Here, the notations $G_{i}$, $G'_{i}$, $T_{i}(z')$ and $T'_{i}(z')$ are the same as in Equation \ref{eq:convencc}. Additionally, $\delta G_{i}$ indicates the uncertainty in $G_{i}$. The mask spectrum, $M_{i}$, assigns a value of one for pixels included in the cross-correlation and zero for those excluded. Bad pixels, sky subtraction residuals, and spectral lines that are not relevant for redshift determination are masked (i.e., $M_i=0$). 

The primary advantage of inverse-variance weighted cross-correlation over the conventional cross-correlation method is the inclusion of the weighting factor $M_{i}/(\delta G_i/G_i')^2$, which accounts for uncertainties in the observed spectrum. Because the weighting factor is higher for pixels with a higher signal-to-noise ratio (S/N), these higher S/N pixels contribute more to the cross-correlation signals. As a result, the inverse-variance weighted cross-correlation effectively filters out noisy features, improving the robustness of the measurement (Kelson et al. 2003).

Another advantage of using the inverse-variance weighted cross-correlation is the straightforward determination of redshift uncertainty (Kelson et al. 2003). Suppose we fit the template spectrum $T$ to the observed spectrum $G$ at redshift $z$ by shifting $T$ along the redshift dimension $z'$. The $\chi^2$ used to optimize the fit is defined as:
\begin{equation}
    \chi^2(z') = \sum^n_{i=1}\frac{M_i}{\delta G^2_i}\left(G_i-T^\circ_i(z')\right)^2.
\end{equation}
Because $T^\circ\approx G'$ near $z'=z$,
\begin{equation}
    \left.\frac{d\chi^2}{dz'}\right|_{z'=z}=-2\sum^n_{i=1}\frac{M_i}{\delta G^2_i}(G_i-T^\circ_i)\left.\frac{dT^\circ_i}{dz'}\right|_{z'=z}\approx-2\sum^n_{i=1}\frac{M_i}{\delta G^2_i}(G_i-G'_i)\left.\frac{dT^\circ_i}{dz'}\right|_{z'=z}=-2\left.\frac{d C.C.}{dz'}\right|_{z'=z},
\end{equation}
thus, $\left.\Delta \chi^2\right|_{z'=z}=-2\left.\Delta C.C.\right|_{z'=z}$. By definition, the $1\sigma$ uncertainty of the best-fit redshift corresponds to the $\Delta \chi^{2}$ centered around the minimum $\chi^{2}$. Therefore, the $1\sigma$ uncertainty of the redshift derived from the cross-correlation signals ($(\delta z)_{\rm c.c.}$) is determined based on:
\begin{equation} \label{eq:uncer}
    C.C.(z\pm(\delta z)_{\rm c.c.}) = C.C.(z)-\frac{1}{2}.
\end{equation}

\subsection{Estimating uncertainty with Gaussian fitting} \label{sec:uncertainty}

We next describe the mathematical procedure for estimating redshift uncertainty based on Equation \ref{eq:uncer}. Assuming the Gaussian shapes of the spectral lines, the cross-correlation signal between two Gaussian lines also follows the Gaussian profile \citep{Tonry79}. Thus, when the observed and shifted template spectra are well-matched (i.e., $z' \approx z$), the expected cross-correlation signals show a Gaussian shape: 
\begin{equation} \label{eq:gaussian}
    C.C.(z')\sim h_{\rm c.c.}e^{-(z'-z)^2/2\sigma_{\rm c.c.}}.
\end{equation}
Then, the solution for Equation \ref{eq:uncer} is:
\begin{equation} \label{eq:sol}
    (\delta z)_{\rm c.c.} + z = \sigma_{\rm c.c.}\sqrt{-2\ln{(1-1/2h_{\rm c.c.}))}}+z.
\end{equation}
In practice, we derive the best-fit Gaussian parameters in Equation \ref{eq:gaussian} from the cross-correlation signal. The best-fit parameters $h_{\rm c.c.}$, $z$, and $\sigma_{\rm c.c.}$ have uncertainties from the Gaussian fit ($(\delta h_{\rm c.c.})_{\rm fit}$, $(\delta z)_{\rm fit}$, and $(\delta \sigma_{\rm c.c.})_{\rm fit}$). Thus, $(\delta z)_{\rm c.c.}$ computed by Equation \ref{eq:sol} is also affected by the fitting uncertainty: 
\begin{equation} \label{eq:sigfit}
    (\delta (\delta z)_{\rm c.c.})_{\rm fit}  = \sqrt{\left[(\delta z)_{\rm fit}+2\sigma_{\rm c.c.}\sqrt{-2/\ln{(1-1/2h_{\rm c.c.})}}\frac{(\delta h_{\rm c.c.})_{\rm fit}}{h_{\rm c.c.}^2(1-1/2)}+2(\delta\sigma_{\rm c.c.})_{\rm fit}\sqrt{-2\ln{(1-1/2h_{\rm c.c.})}}\right]^2+(\delta z)_{\rm fit}^2}. 
\end{equation}

Finally, we define the redshift uncertainty by taking into account the uncertainty from the Gaussian fitting (i.e., $(\delta (\delta z)_{\rm c.c.})_{\rm fit}$):
\begin{multline} \label{eq:sigz}
    \delta z = (\delta z)_{\rm c.c.} + (\delta (\delta z)_{\rm c.c.})_{\rm fit} =\sigma_{\rm c.c.}\sqrt{-2\ln{(1-1/2h_{\rm c.c.}))}}
        \\+ \sqrt{\left[(\delta z)_{\rm fit}+2\sigma_{\rm c.c.}\sqrt{-2/\ln{(1-1/2h_{\rm c.c.})}}\frac{(\delta h_{\rm c.c.})_{\rm fit}}{h_{\rm c.c.}^2(1-1/2)}+2(\delta\sigma_{\rm c.c.})_{\rm fit}\sqrt{-2\ln{(1-1/2h_{\rm c.c.})}}\right]^2+(\delta z)_{\rm fit}^2}.
\end{multline}

\subsection{Reliability of Redshift Measurements} \label{sec:significance}

We introduce two metrics to assess the reliability of redshift measurements obtained using the cross-correlation technique: $\chi^2_{\rm eff}$ and $r-$value. These parameters are useful not only for determining the significance of the cross-correlation signal but also for evaluating the reliability of redshift measurements. 

$\chi^2\mathrm{_{eff}}$ measures the level of consistency between an observed spectrum and a template spectrum. $\chi^2\mathrm{_{eff}}$ is defined as follows:
\begin{equation}
    \chi^2\mathrm{_{eff}} = \sum^n_{i=1}\frac{M_i}{\delta G^2_i}\left[G_i-T^\circ_i(z')\right]^2/\left[\left(\sum^n_{i=1}M_i\right)-n\mathrm{_{conti}}-1]\right].
    \label{eq:chi_eff}
\end{equation}

Here, $\frac{M_i}{\delta G^2_i}\left[G_i-T^\circ_i\right]^2$ is $\chi^2$ and $\left[\left(\sum^n_{i=1}M_i\right)-n\mathrm{_{conti}}-1]\right]$ is a degree of freedom where $n_{\rm conti}$ is the number of constraints determined by the procedure of tracing continuum (see Section \ref{sec:process} and \ref{sec:z_from_cc}). $\chi^2\mathrm{_{eff}}$ becomes closer to one as the template spectrum matches the observed spectrum. We consider $\chi^2_{\rm eff}<4$ means that the template spectrum matches the observed spectrum well because $\chi^2_{\rm eff}=4$ roughly corresponds to $\sim 2\sigma$ difference.

Another parameter, the $r$-value, was introduced by \citet{Tonry79}. The $r$-value quantifies the significance of a cross-correlation peak and is defined as follows \citep{Tonry79}:
\begin{equation}
    \text{r}=h/\sigma_a,
\end{equation}
where $h$ is the height of the peak in the cross-correlation signal and $\sigma_a$ is computed as
\begin{equation}
    \sigma^2_a = \frac{1}{2N}\sum^{m=N}_{m=1}[\text{C.C.}(n_z-m)-\text{C.C.}(n_z+m)]^2.
\end{equation}
Here, $C.C(n_z\pm m)$ represents the cross-correlation signal at $n_z \pm m$-th redshift lag, where $n_z$-th lag corresponds to $z'=z$. We use the number of redshift lags within $(z-0.1,~z+0.1)$ as $N$, which is large enough to include the peak and the adjacent noisy features of the cross-correlation signals. The $r$-value indicates the significance of the true peak relative to the noise, effectively serving as the signal-to-noise ratio of the cross-correlation signal \citep{Tonry79}. A higher $r$-value generally indicates a more reliable redshift measurement. In general, $r > 3 - 5$ suggests that the cross-correlation signal is significant \citep{Kurtz98, Geller14, Geller16}.

\section{RVSNUpy} \label{sec:rvsnupy}
In this section, we explain how our Python package, \texttt{RVSNUpy} \citep{RVSNUpy} \footnote{The source code is also available here: https://github.com/kim8517/RVSNUpy}, performs redshift measurements based on the cross-correlation technique. We use Python, a language widely used in and readily accessible to the astronomical community. To improve computational efficiency, we introduce parallel computation based on \texttt{joblib}\footnote{https://github.com/joblib/joblib} and minimize the use of serial computation to reduce the computation time.

\begin{figure}[h]
    \centering
    \includegraphics[width=.8\textwidth]{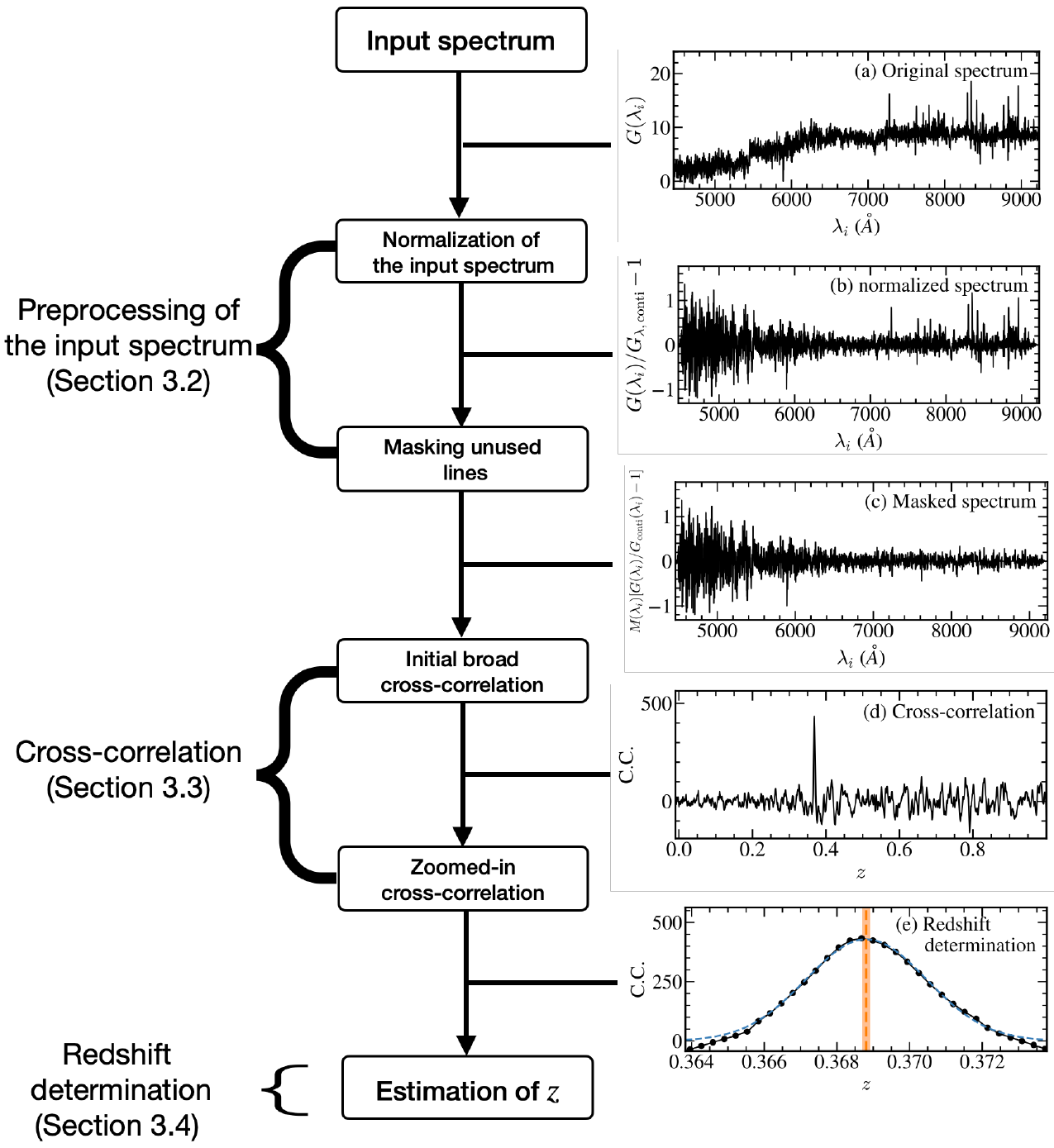}
    \caption{The flow chart illustrating how \texttt{RVSNUpy} works. The plots on the right side illustrate the resulting products as \texttt{RVSNUpy} proceeds. (a) An example input spectrum obtained by SDSS. (b) The input spectrum normalized with the continuum derived by B-spline fit. (c) The normalized input spectrum after masking emission lines and bad pixels. (d) The cross-correlation signal between the input spectrum and the template spectrum as a function of redshift lag within the full redshift range of $z < 1.$. (e) Redshift determination based on the cross-correlation signals. Black circles and solid lines show the cross-correlation signals. The blue-dashed line displays the best-fit Gaussian. The orange vertical line indicates our final redshift estimation, corresponding to the mean of the best-fit Gaussian function. The orange-shaded region marks the uncertainty of the redshift.}  
    \label{fig1}
\end{figure}

Figure \ref{fig1} illustrates the overall structure of \texttt{RVSNUpy}. First, we prepare a set of template spectra for the cross-correlation (Section \ref{sec:templates}). We normalize and mask the input spectrum to derive the cross-correlation signal based only on spectral lines (Section \ref{sec:process}). \texttt{RVSNUpy} then computes the cross-correlation signal by shifting the template spectrum across a wide range of radial velocities (Section \ref{sec:cross}). \texttt{RVSNUpy} measures the redshift of the input spectrum by identifying the peak of the cross-correlation signal (Section \ref{sec:z_from_cc}). Finally, \texttt{RVSNUpy} determines the final redshift among the cross-correlation results with various template spectra (Section \ref{sec:determination}). A detailed, step-by-step explanation of the procedure is provided in the following sections. 

\subsection{Template spectra} \label{sec:templates}

Template spectra are an essential ingredient of redshift measurement. \texttt{RVSNUpy} requires the template spectra to satisfy three key criteria. First, the template spectra must be in the rest frame (i.e., $z = 0$). A non-zero radial velocity of the template spectra introduces bias in the redshift determination during cross-correlation. Second, the template spectra must be uniformly sampled on a logarithmic wavelength scale. This ensures that shifting the template spectra along logarithmically spaced wavelength pixels corresponds to a shift in redshift space (see Section \ref{sec:cc}). If the template spectra are not logarithmically sampled, the mathematical framework outlined in \ref{sec:cc} becomes invalid.

Lastly, the template spectra must have a significant signal-to-noise ratio and a small pixel scale. The redshift determination described in Section \ref{sec:cc} assumes that the input spectrum and template spectra have identical pixel scales. However, in practice, spectra obtained with different spectrographs often differ in pixel scale. \texttt{RVSNUpy} interpolates the template spectrum to match the pixel scale of the input spectrum. The interpolation introduces slight deviations from the original template spectrum due to its inherent limitations. Therefore, to minimize the loss of spectral information during the interpolation, \texttt{RVSNUpy} requires that the template spectra have significant signal-to-noise ratios and small pixel scales.

\texttt{RVSNUpy} uses a set of multiple template spectra, including characteristic spectra of various types of objects. Because the input spectrum may contain different absorption and emission lines, relying on a single template spectrum can create large uncertainties and unreliable redshift measurements. To mitigate this, our template set includes spectra of various astronomical objects. 

Additionally, RVSNUpy includes a keyword that indicates whether the template spectra are dominated by absorption or emission lines. The user needs to manually specify this keyword for \textit{each template spectrum (rather than input spectrum)} in \texttt{RVSNUpy}. This indication is crucial for reliable redshift determination, particularly if the input spectrum contains both absorption and emission lines. Absorption lines typically originate from stellar components, whereas emission lines originate from gaseous regions. The redshifts derived from two different types of lines may differ. If the redshift of the object is measured based on the different types of lines, the resulting redshift measurement may be unreliable. Additionally, cross-correlation that occurs between absorption lines in the input spectrum and the emission lines in the template spectrum (or vice versa) can weaken the cross-correlation signal, further compromising the accuracy of the redshift determination. To derive the redshift based on the cross-correlation between similar types of spectral lines, the template spectrum should be classified as either absorption-line-dominated (hereafter absorption line template) or emission-line-dominated (hereafter emission line template). Based on the designated class of template spectrum, \texttt{RVSNUpy} selects the corresponding spectral lines in the input spectrum for performing cross-correlation (see Section \ref{sec:process}).

\subsection{Preprocessing of Input Spectra} \label{sec:process}

\texttt{RVSNUpy} preprocesses the input spectra prior to performing the cross-correlation. The preprocessing involves two main steps: normalizing the input spectrum and masking spectral lines or artifacts that could interfere with accurate redshift determination. Panels (a) to (c) in Figure \ref{fig1} illustrate the preprocessing steps applied to an example SDSS spectrum. Figure \ref{fig1} (a) shows an example SDSS spectrum we use for this demonstration. The input spectrum consists of the wavelength ($\lambda_i$), fluxes ($G_{i}$), flux uncertainties ($\delta G_{i}$), and the mask ($M^{0}_{i}$). The mask $M^{0}_{i}$ indicates bad pixels (i.e., mask values equal to zero) associated with unreliable fluxes caused by saturation, cosmic rays, or poorly subtracted sky lines. During preprocessing, \texttt{RVSNUpy} first normalizes the spectrum by its continuum (Figure \ref{fig1} (b)) and then masks the normalized spectrum to exclude spectral lines not used for cross-correlation (Figure \ref{fig1} (c)). 

The first preprocessing step is continuum normalization. Cross-correlating the full spectrum, including the continuum and spectral lines, with template spectra artificially enhances the cross-correlation signal due to the non-zero continuum flux, leading to inaccurate redshift measurements (see Section \ref{sec:cc}). To address this, \texttt{RVSNUpy} removes the continuum to improve the precision of the redshift determination. 

To trace the continuum, \texttt{RVSNUpy} fits a B-spline function \citep{deBoor78, Dierckx95} to the input spectrum. We use a knot span of $100~\mathrm{\AA}$, which is narrow enough to follow the shape of the continuum but broad enough to prevent the fit from tracing small features like spectral lines. When more than 75\% of pixels within a node (a single knot span) are masked, \texttt{RVSNUpy} merges the node with the adjacent one at longer wavelengths to derive a reliable fit. 

Because spectral lines can contaminate the B-spline continuum fit, \texttt{RVSNUpy} identifies and masks these spectral lines in three steps. First, the initial best-fit B-spline model is subtracted from the input spectrum. Second, regions where the residuals exceed twice the standard deviation of the residual fluxes are flagged. Third, \texttt{RVSNUpy} derives the best-fit Gaussian for the residual spectra in the flagged regions. If the Full Width at Half Maximum (FWHM) of the best-fit Gaussian exceeds the spectral resolution of the input spectrum, the region is marked as containing the spectral lines. 

After masking the marked spectral lines, \texttt{RVSNUpy} refits the continuum with inverse-variance weighting for each pixel ($M^0_i / \delta G_i^2$). The input spectrum is then divided by the refined continuum ($G_i / G'_i$) and shifted to zero by subtracting one, as described in Equation \ref{eq:wcc}. To ensure smooth transitions at the spectrum’s boundaries, \texttt{RVSNUpy} applies a Tukey window function \citep{Tukey67} to the normalized spectrum. Figure \ref{fig1}(b) shows the resulting normalized spectrum, where the continuum has been removed, leaving only spectral lines and noise, which are critical for determining the redshift and its uncertainty.

The second preprocessing step involves generating a mask spectrum for computing inverse-variance weighted cross-correlation signals, which we will refer to (and was mentioned in Equation \ref{eq:wcc}) as $M_i$, to indicate which spectral lines are used. This mask combines the initial bad pixel mask ($M^{0}_{i}$) with additional markings for spectral lines. \texttt{RVSNUpy} uses spectral lines identified during the continuum fitting process. Spectral lines with minimum fluxes below zero are classified as absorption lines, while those with maximum fluxes above zero are classified as emission lines. The mask spectrum ($M_i$) is annotated to indicate the positions of these absorption and emission lines, which are subsequently used in the cross-correlation with the template spectrum (see Section \ref{sec:cross}).

Figure \ref{fig1} (c) shows an example preprocessed spectrum for cross-correlation with the absorption line template. The input spectrum is normalized to the continuum. Additionally, bad pixels, cosmic rays, sky residuals, and emission lines in the spectrum are masked: i.e., $M_i(G_i/G'_i)$ in Equation \ref{eq:wcc}. 

\subsection{Cross-correlation} \label{sec:cross}

\texttt{RVSNUpy} computes the inverse-variance weighted cross-correlation signals based on Equation \ref{eq:wcc} by shifting the template spectrum along the wavelength direction. The user can adjust the starting shift for the cross-correlation. Then, \texttt{RVSNUpy} iteratively computes the cross-correlation signal as it shifts the template spectrum. Because \texttt{RVSNUpy} operates with the template spectrum on a logarithmic wavelength scale, the cross-correlation signal computed as the template spectrum shifts (pixel-by-pixel) is a function of the redshift lag $z'$. This pixel shift corresponds to redshift ($\Delta z'=\Delta \log{\lambda_i}$) and the velocity lag ($\Delta v'=c\Delta z'$). 

Before we start the cross-correlation, we match the wavelength systems for templates and input spectra between the air and vacuum wavelength systems. The vacuum and air wavelength systems indicate whether the wavelength of a spectrum is calibrated in vacuum or air. The conversion between two systems is non-negligible; the vacuum wavelength is slightly redshifted by $\sim 1.7~{\rm \AA}$ at $\sim 6000~{\rm \AA}$ (corresponding to $\sim 85~\kms$). The mismatch of the wavelength system causes $\sim 50 - 100~\kms$ systematic redshift offsets in the redshift measurements based on optical spectra.

RVSNUpy enables redshift derivation in both vacuum and air wavelength systems. The users can specify which wavelength system they want to use for redshift measurement. Once the wavelength system is specified, RVSNUpy converts the wavelength system based on the following equation \footnote{https://sdss.org/dr18/tutorials/conversions/} \citep{Ciddor96}:
\begin{equation}
    \lambda_{\mathrm{air}} = \frac{\lambda_{\mathrm{vac}}}{
    1.0 + 5.792 \times 10^{-2}/[238.019 - (10^4/\lambda_{\mathrm{vac}})^2]
    + 1.679 \times 10^{-3}/[57.362 - (10^4/{\lambda_{\mathrm{vac}}})^2]},
\end{equation}
where $\lambda_{\mathrm{air}}$ and $\lambda_{\mathrm{vac}}$ are air and vacuum wavelengths in \AA, respectively. We note that we used the vacuum wavelength system for tests with the synthetic spectra in Section \ref{sec:modeltest} and we used the air wavelength system for tests with the observed spectra in Section \ref{sec:obstest}.

\texttt{RVSNUpy} manipulates the template spectrum in two ways to cross-correlate with the input spectrum. First, the template spectrum is interpolated to match the pixel scale of the input spectrum, which is often linearly scaled. To compute the cross-correlation lag, both the input spectrum and template spectrum should be on the same scale. At each redshift lag (corresponding to a pixel shift)  \texttt{RVSNUpy} resamples the template spectrum to match the wavelength scale of the input spectrum. Resampling the template spectrum with high S/N and small pixel scale retains more spectral information than interpolating the input spectrum.  

Second, \texttt{RVSNUpy} removes the continuum from the template spectrum using the same procedure applied to the input spectrum (Section \ref{sec:process}). \texttt{RVSNUpy} first derives an initial continuum based on the B-spline function fitted to the template spectrum. Then, the spectral lines are identified after subtracting the initial continuum from the template spectrum. After masking the spectral lines, we derive the best-fit B-spline function as the final continuum of the template spectrum. For continuum removal, \texttt{RVSNUpy} uses the same weight function previously used to determine the continuum of the input spectrum, ensuring equal weighting of the input spectrum and template spectrum for each pixel. Because we use the weight function based on the fixed input spectrum, \texttt{RVSNUpy} needs to determine the continuum of the template spectrum at each pixel shift. 

The repeated interpolation and continuum determination of the template spectrum at each pixel shift requires a significant computation time. For effective computation, \texttt{RVSNUpy} implements a two-step approach. The first step is an initial search for a plausible cross-correlation peak. The second step is the zoomed-in cross-correlation around the cross-correlation peak identified in the first step. This two-step approach significantly reduces the computation time by an order of magnitude compared to determining the continuum for each pixel shift. 


During the initial search for a plausible cross-correlation peak, \texttt{RVSNUpy} removes the continuum from the template spectrum without applying the weight function. In this step, the input spectrum is also interpolated to rescale them onto a logarithmic wavelength scale. Although interpolation can result in some loss of information, we employ this process to reduce the computation time. The logarithmically interpolated input spectrum is then cross-correlated with the continuum-removed template spectrum, yielding a preliminary cross-correlation result with reduced computation time. \texttt{RVSNUpy} identifies a peak (or peaks) as any value exceeding half of the maximum cross-correlation signal or the standard deviation of the cross-correlation signal derived from the preliminary cross-correlation result. 

In the zoomed-in cross-correlation, \texttt{RVSNUpy} performs interpolation and continuum removal on the template spectrum at each redshift shift in the identified peaks. In this step, \texttt{RVSNUpy} uses the same weight as the input spectrum to determine the continuum. \texttt{RVSNUpy} then computes the precise cross-correlation signals based on the interpolated and continuum-removed template spectrum. 

Figure \ref{fig1} displays (d) the cross-correlation result as a function of the velocity lag (i.e., redshift). The cross-correlation signals show a strong peak at $z\sim0.36$, corresponding to the predicted redshift of the input spectrum. In the next step. \texttt{RVSNUpy} derives the best redshift estimate based on the zoomed-in cross-correlation signals.

\subsection{Redshift derivation based on the cross-correlation signals}\label{sec:z_from_cc}

The zoomed-in cross-correlation yields a cross-correlation signal as a function of redshift lag. \texttt{RVSNUpy} determines the redshift by identifying the peak of the cross-correlation signals through Gaussian fitting. The mean of the best-fit Gaussian corresponds to the redshift determined based on the cross-correlation between the observed spectrum and template spectrum.

Figure \ref{fig1} illustrates (e) the redshift determination of \texttt{RVSNUpy}. The black points and solid lines show the zoomed-in cross-correlation signals, and the blue dashed line displays the best-fit Gaussian. The orange vertical line indicates the mean of the best-fit Gaussian, corresponding to the redshift derived from the cross-correlation signal.

The redshift determination is straightforward when a single peak appears in the cross-correlation signal, which occurs in most cases. There are some cases where multiple peaks appear in the cross-correlation signal. For these cases, we use a different approach to determine the redshift depending on the type of template spectrum used to compute the cross-correlation signal. 

If multiple peaks arise from cross-correlation with the absorption line template, \texttt{RVSNUpy} uses $\chi^2_{\rm eff}$ (see Section \ref{sec:significance}) to determine the redshift. Here, $n_{\rm conti}$ in Equation \ref{eq:chi_eff} is the number of knots used for the B-spline fit to the template spectrum. A lower $\chi^2_{\rm eff}$ indicates a better match between the input spectrum and the shifted template spectrum. Thus, the redshift determined from the cross-correlation peak with the lowest $\chi^{2}_{\rm eff}$ is the best prediction. 

In cases involving multiple peaks from cross-correlation with the emission line template, \texttt{RVSNUpy} requires an additional process. Because variations in the strength of emission lines are much larger than absorption lines, the $\chi^{2}_{\rm eff}$ reflecting the resemblance between the input spectrum and template spectrum varies significantly depending on the emission line strength. In other words, a strong variation in $\chi^{2}_{\rm eff}$ (computed with emission line templates) prevents the determination of which cross-correlation peak represents the true redshift of the input spectrum. 

We thus implement an alternative empirical technique to derive the redshift in the particular cases where multiple peaks occur in emission line template cross-correlation. For each peak in the cross-correlation signal, \texttt{RVSNUpy} computes the corresponding redshift ($z'$) based on Gaussian fitting. For each redshift, \texttt{RSNUpy} generates a simple normalized emission line model spectrum ($T^{\rm EM.}$) consisting of the zero continuum and the Gaussian-like emission lines shifted by $z'$. The peak and width of the Gaussian functions are derived from the Gaussian fit to the normalized input spectrum over the wavelength range by assuming that emission lines would appear at $\lambda' = \lambda_{o} (1 + z')$ ($\lambda_{o}$ indicates the rest-frame wavelength of the lines). In some cases, the forced Gaussian fit at $\lambda'$ is unreliable. In these cases, either the height of the peak is below zero or the FWHM of the fit is smaller than the spectral resolution of the spectrograph. We exclude these lines when building the model spectrum. After generating the model spectrum at each $z'$, \texttt{RVSNUpy} computes the $\chi^2_{\rm EM. lines}$ defined as:
\begin{equation}
\label{eq:chi_em}
    \chi^2_{\rm EM. lines} = \sum^N_i\frac{M_i}{(\delta G_i/G'_i)^2}\left(\frac{G_i}{G'_i}-T^{\rm EM.}_i\right)^2/\left[\left(\sum^n_{i=1}M_i\right)-3n_{\rm EM. line}-1]\right].
\end{equation}
Here, $n_{\rm EM. line}$ indicates the number of emission lines.
The main difference between $\chi^{2}_{\rm EM. lines}$ and $\chi^{2}_{\rm eff}$ is that the $\chi^{2}_{\rm EM. lines}$ is computed based on the normalized input and emission-line model spectrum, while $\chi^{2}_{\rm eff}$ is calculated based on the input and the template spectra before normalizations. \texttt{RVSNUpy} finally determines the redshift as the peak with the lowest $\chi^{2}_{\rm EM. lines}$.

\subsection{Final redshift determination from various template spectra} \label{sec:determination}
\texttt{RVSNUpy} iteratively applies the whole process described in Section \ref{sec:process}, \ref{sec:cross} and \ref{sec:z_from_cc} to derive redshifts for a set of template spectra which includes various types of objects (e.g., stars, absorption- or emission-line dominated galaxies, quasars). 

\texttt{RVSNUpy} determines the final redshift of the input spectrum by quantifying the quality of redshift measurements based on $r-$value and $\chi^2_{\rm eff}$ (or $\chi^2_{\rm EM. lines}$). \texttt{RVSNUpy} first excludes measurements with $r < r_{\rm thres}$ and $\chi^2_{\rm eff}~({\rm or}~ \chi^2_{\rm EM. lines})>\chi^2_{\rm thres}$. Reliable values of these parameters range between $3 < r_{\text{thres}} < 5$ and $2 < \chi^2_{\text{thres}} < 4$, respectively. If a single redshift measurement satisfies these selections, \texttt{RVSNUpy} returns that value.

In the cases of multiple redshift estimates that satisfy the $r_{\rm thres}$ and $\chi^{2}_{\rm thres}$ requirements, we choose the final redshift measurements with the following steps. First, the redshift measurements based on absorption line templates are prioritized over those based on emission line templates. We adopt this selection because the light from galaxies (i.e., our primary targets) is dominated by the stellar components with absorption lines. If there are multiple measurements for the absorption line templates, the measurement with the lowest $\chi^2_{\rm eff}$ is chosen as the final redshift. Second, if there is no redshift measurement based on the absorption line templates satisfying the $r_{\rm thres}$ and $\chi^{2}_{\rm thres}$ selections, \texttt{RVSNUpy} determines the measurement based on the emission line template with the lowest $\chi^{2}_{\rm EM, lines}$ as the final redshift. In general, \texttt{RVSNUpy} measures the redshift for a single spectrum in 0.2-0.4 seconds. The final products of \texttt{RVSNUpy} include the final redshift measurement of the input spectrum and redshift estimations based on multiple templates along with their $\chi^2_{\rm eff}$ and $r-$values. 

\section{Sets of template spectra}

Here, we describe an example set of template spectra to test the performance of \texttt{RVSNUpy}. We note that users can construct their own template spectra for their own purposes given that the template spectra satisfy the requirements described in Section \ref{sec:templates}.

The first set of template spectra we use is a set of SDSS cross-correlation template spectra \footnote{SDSS uses the cross-correlation technique for redshift measurements up until Data Release 6 \citep{sdss_dr6}}. There are 33 template spectra obtained with the SDSS spectrograph, including various types of stars (23), galaxies (6), and quasars (4) \footnote{https://classic.sdss.org/dr5/algorithms/spectemplates/}. Because our main interest is measuring the redshift of galaxies, we select three out of six galaxy template spectra in the SDSS template set: `Early-type galaxy', `Luminous red galaxy', and `Late-type galaxy'. The `Early-type galaxy' and `Luminous red galaxy' are absorption line templates, and the `Late-type galaxy' is an emission line template. These three galaxy template spectra reflect the most characteristic types of galaxies. We refer to these three template spectra as the `SDSS templates'. 

Another set of template spectra we use is the template spectra set for measuring the redshifts of spectra obtained with MMT/Hectospec \citep{Kurtz98}. These template spectra are empirically constructed from Hectospec observations and are named: `eatemp', `eltemp', `habtemp90', `sptemp', and `hemtemp0.0'. The `eatemp', `eltemp', and `habtemp90' are absorption line templates and the `sptemp', and `hemtemp0.0' are emission line templates. We refer to these five template spectra as the `Hectospec templates'.

We also construct a set of template spectra based on synthetic spectra. We generate the synthetic spectra from single stellar population (SSP) models with various ages (0.01, 3, 5, 7, 9, 11 Gyr) using the stellar population synthesis code (Flexible Stellar Population Synthesis, \texttt{FSPS}; \citealp{Conroy09, Conroy10}). We use \texttt{FSPS} because it enables the construction of SSP spectra with emission lines. For example, the SSP spectrum we generate with \texttt{FSPS} with an age of 0.01 Gyr shows multiple strong emission lines (e.g., H$\beta$, [O III], H$\alpha$). We use this 0.01 Gyr spectrum as the emission template and the spectra with older stellar ages as absorption templates. We refer to these synthetic template spectra as the `\texttt{FSPS} templates'. 

\section{Tests with synthetic spectra} \label{sec:modeltest}

We test the performance of \texttt{RVSNUpy} using the synthetic spectra generated from \texttt{FSPS} \citep{Conroy09, Conroy10}. By varying S/N, resolution, and the shape of spectra, the tests with model spectra allow for the investigation of potential systematic biases of \texttt{RVSNUpy}.

\begin{figure}[h]
    \centering
    \includegraphics[width=1.1\textwidth]{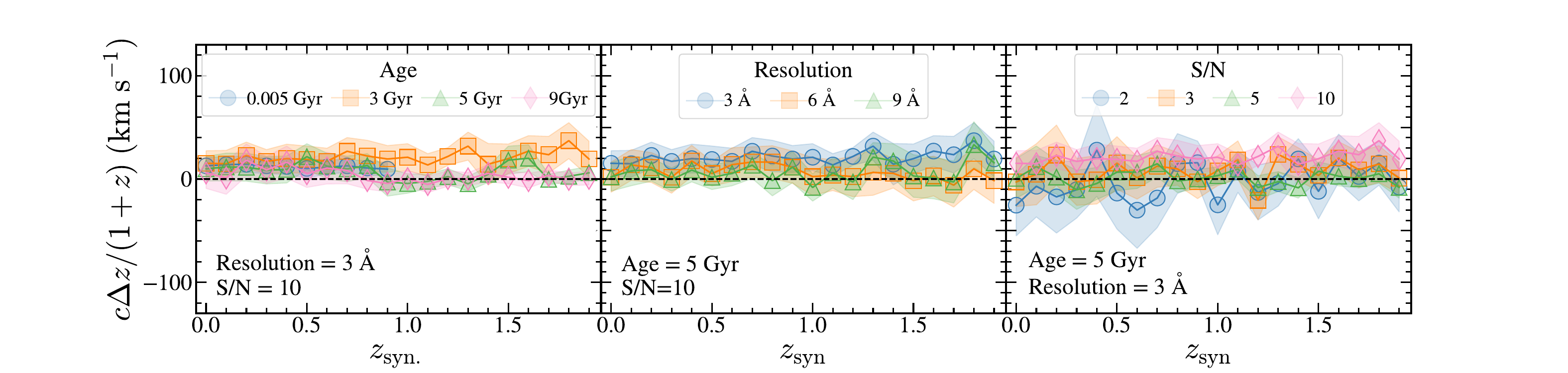}
    \caption{Comparison between \texttt{RVSNUpy} redshift estimates and true redshifts for a set of synthetic SSP spectra. The left panel shows the deviation of \texttt{RVSNUpy} redshift with respect to the redshift of the synthetic spectra ($c\Delta z/(1+z)=c(z^{\rm RVSNUpy} - z_{\rm syn}) / (1 + z_{\rm syn})$) with various stellar population ages, but with fixed resolution and S/N. The middle and right panels display similar comparisons but with synthetic spectra with various resolutions and S/Ns. In each panel, other parameters used for the comparison are fixed. The shaded regions with various colors display the $1\sigma$ uncertainty of the redshift difference for comparisons with various synthetic spectra.}
    \label{fig2}
\end{figure}

We first generate three single stellar population (SSP) spectra with ages of 0.005, 3, 5, and 9 Gyrs with a fixed spectral resolution of $3 {~\rm\AA}$ and S/N of 10. The 0.005 Gyr spectrum is emission line dominated, and others are absorption line dominated. This set of spectra with different population ages enables performance testing of \texttt{RVSNUpy} depending on the spectral shape. 

Similarly, we generate the spectra with various spectral resolutions ($3~\rm \AA$, $6~\rm \AA$, and $9~\rm \AA$) with a fixed stellar population age of 5 Gyr and S/N of 10. The initial \texttt{FSPS}-synthesized SSP spectra have a resolution of $3~{\rm \AA}$. The required broadening to achieve broader spectral resolutions ($\sigma_{\rm target}$) is $\sigma_{\rm broadening} = \sqrt{\sigma_{\rm target}^2 - (3~{\rm \AA})^{2}}$. We apply the Gaussian convolution with the kernel size of  $\sigma_{\rm broadening}$ to the original SSP spectrum to generate two additional synthetic spectra with spectral resolutions of $6~\rm \AA$ and $9~\rm \AA$. 

We also add different amounts of noise to generate synthetic spectra with different S/N (2, 3, 5, and 10). For this set of spectra, we fix the stellar population age at 5 Gyr and the spectral resolution at 3 \AA. We first compute a normalized factor ($\alpha$) that yields an expected S/N for the median flux. We assume the noise of the spectrum ($\delta G_0$) is Poisson. For the expected $S/N$, the normalized factor is defined as follows:
\begin{equation}
S/N = \frac{G\mathrm{_{0, med}}}{\delta G\mathrm{_{0, med}}} = \frac{\sqrt{G\mathrm{_{0,med}}}}{\alpha}
\end{equation}
Based on $\alpha$, we compute the tentative range of flux uncertainty at each wavelength: $\delta G_i=\alpha\sqrt{G_{0,i}}$. The final spectrum extracted from these flux uncertainty ranges has the expected S/N. Thus, we determine the final flux at each wavelength by randomly sampling based on the Gaussian with the mean as the flux of the original synthetic spectra and $1\sigma$ as the computed flux uncertainty. 

For our test, we shift the synthetic spectra to different redshifts between $z = 0-1$ for 0.005 Gyr spectra and $z= 0-2$ for other spectra in intervals of 0.1. \footnote{We limit the test redshift range because the SDSS templates cover limited wavelength ranges. The measurable redshift range can be extended to higher redshifts once templates covering wider wavelength ranges are prepared.} The final test set includes combinations of three different stellar population ages, spectral resolutions, and S/N, as well as ten or twenty different redshifts. We measure the redshift of the synthetic SSP spectra based on \texttt{RVSNUpy} with the SDSS templates. Here, we use the vacuum wavelength system for the synthetic spectra and the SDSS templates.

Figure \ref{fig2} shows the difference between redshifts measured by \texttt{RVSNUpy} ($z^{\rm RVSNUpy}$) and the redshifts of the synthetic SSP spectra ($z_{\rm syn}$): $c(z^{\rm RVSNUpy} - z_{\rm syn}) / (1 + z_{\rm syn})$ as a function of $z_{\rm syn}$. The left panel of Figure \ref{fig2} displays the test results based on synthetic spectra with various stellar population ages while keeping spectral resolution and S/N fixed. Similarly, the middle and right panels show the results based on the synthetic SSP spectra with different spectral resolutions (middle) or S/N (right), with all other parameters fixed. 

\texttt{RVSNUpy} generally derives consistent redshifts across all redshift ranges of the synthetic spectra. The typical redshift uncertainty is $\sim 20~\kms$, and the redshift differences are generally within $1\sigma$. The redshift measurements show small systematic offsets ($\sim 15~\kms$), but remain within $1\sigma$ uncertainties. The offsets originate from the radial velocity calibration of the SDSS templates (see Section \ref{sec:offsetorigin}). We will discuss the calibration of the templates to resolve the systematic offset in the redshift measurement in Sections \ref{sec:offsetorigin} and \ref{sec:obsconclusion}.

The redshift uncertainty does not significantly depend on the spectral resolutions of the synthetic SSP spectra, but increases as S/N decreases. The dependence of redshift uncertainty on spectral resolutions and S/N is related to the accuracy in determining the center of the spectral lines. Ideally, spectral line shapes follow a Gaussian model \citep{Tonry79}. The mean of the Gaussian is determined accurately regardless of the width of the Gaussian as long as the pixel scale of the spectrum satisfies the Nyquist condition \citep{Shannon49}. Thus, the location of the spectral lines can be measured accurately and is only weakly dependent on the spectral resolution. On the other hand, in the case of noisy spectra, the shape of the spectral lines deviates from Gaussian. In this case, locating spectral lines becomes more difficult, which leads to greater redshift uncertainty.

The tests with synthetic spectra show that \texttt{RVSNUpy} consistently derives redshifts from the synthetic spectra. The derived redshifts are identical to the input redshift within the uncertainty range for various ages, resolutions, and S/N. The consistency between the derived redshifts and the input redshifts confirms that \texttt{RVSNUpy} successfully recovers redshifts regardless of spectral properties.

\section{Tests with Observational Data} \label{sec:obstest}

Next, we tested the performance of \texttt{RVSNUpy} using observational spectra obtained with SDSS and MMT/Hectospec spectrographs. We first compared the RVSNUpy redshift measurements based on the SDSS and Hectospec templates with the redshift measurements from the SDSS catalog in Section \ref{sec:obsres}. We describe the impact of template selection on the redshift measurements in Section \ref{sec:offsetorigin}. In Section \ref{sec:obsconclusion}, we discuss the universality of \texttt{RVSNUpy}'s ability to derive consistent redshifts from various types of spectra. 

\subsection{\texttt{RVSNUpy} performance tested with SDSS and Hectospec spectra} \label{sec:obsres}

We derive the redshifts from the observed spectra using \texttt{RVSNUpy}. For this test, we use the HectoMAP redshift survey \citep{Geller11, Hwang16, Sohn21, Sohn23}, a spectroscopic survey conducted with MMT/Hectospec \citep{Fabricant05}. The HectoMAP redshift survey includes $\sim 100,000$ spectra for galaxies within $z \lesssim 0.8$. HectoMAP DR2 \citep{Sohn23} contains 6009 objects observed by SDSS and MMT/Hectospec.  Hereafter, we refer to these 6009 objects as the HectoMAP test sample. The wavelengths of both SDSS and MMT/Hectospec spectra for the HectoMAP test sample are heliocentric corrected. For all input and template spectra in Section \ref{sec:obstest}, we adopt the air wavelength system for a fair comparison between the SDSS and MMT/Hectospec spectra.

We will use the following notation to refer to different types of redshift measurements by specifying the observed spectra, template spectra, and a redshift measurement tool:
\begin{equation}
z\mathrm{_{input~spectra, templates}^{tools}}.
\end{equation}
For example, $z\mathrm{_{Hecto, Hecto}^{RVSNUpy}}$ indicates the redshifts derived from the Hectospec spectra based on \texttt{RVSNUpy} with Hectospec templates.

The HectoMAP test sample also includes the redshifts from SDSS spectra. The redshifts of SDSS spectra are measured with \texttt{Redmonster} \citep{Hutchinson16}, a full spectral fitting tool that uses a galaxy template set based on \texttt{FSPS} \citep{Conroy09, Conroy10} synthetic spectra and a stellar template set based on Kurucz \texttt{ATLAS9} \citep{Meszaros12} and {ASSòT} \citep{Koesterke09}. For simplicity, we refer to these SDSS redshifts as $z\mathrm{^{SDSScat}}$. Table \ref{tab1} summarizes the observed spectra, redshift templates, and the redshift measurement techniques we use. 

\begin{deluxetable*}{lll}
\label{tab1}
\tablecaption{Notation for Redshift Measurements}
\tablecolumns{3}
\small
\tablewidth{0pt}
\tablehead{\colhead{Items} & \colhead{Symbol} & \colhead{Description}}
\startdata
\multirow{2}{*}{Input~spectra} & $z\mathrm{_{SDSS}}$ & The redshifts of SDSS spectra. \\
                               & $z\mathrm{_{Hecto}}$  & The redshifts of Hectospec spectra. \\
\hline
\multirow{1}{*}{Tools}         & $z\mathrm{^{RVSNUpy}}$ & The redshifts measured using \texttt{RVSNUpy} \\
\hline
\multirow{4}{*}{Templates}     & $z\mathrm{_{,SDSS}}$ & The redshifts measured with SDSS templates. \\
                               & $z\mathrm{_{,Hecto}}$  & The redshifts measured with Hectospec templates. \\
                               & $z\mathrm{_{,FSPS}}$ & The redshifts measured with \texttt{FSPS} templates.  \\
                               & $z\mathrm{_{,utemp}}$ & The redshifts measured with our universal templates.  \\
\hline
\multirow{1}{*}{Others}       & $z\mathrm{^{SDSScat}}$ & The redshifts of SDSS spectra from the SDSS catalog.\\
\enddata
\end{deluxetable*}

\begin{figure}[h]
    \centering
    \includegraphics[width=\linewidth]{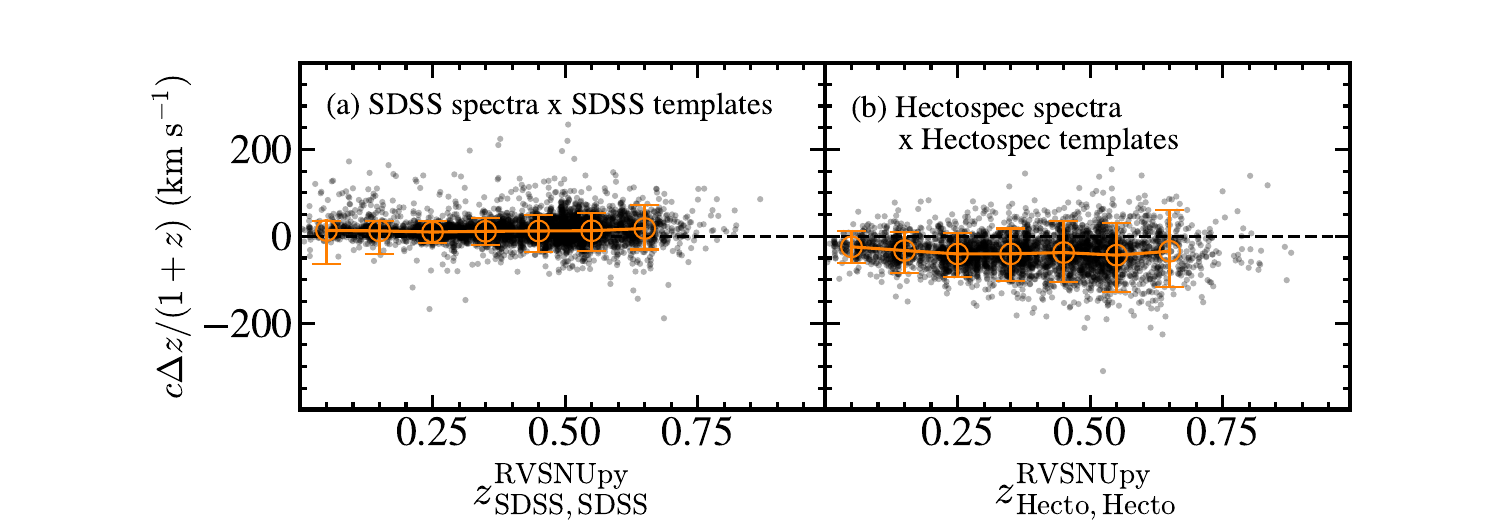}
    \caption{(Left) Comparison between RVSNUpy redshifts measured from SDSS spectra based on the SDSS templates (i.e., $z\mathrm{^{RVSNUpy}_{SDSS, SDSS}}$) and the SDSS redshifts ($z\mathrm{^{SDSScat}}$) for the HectoMAP test sample($c\Delta z /(1+z) = c(\zpysdsssdss-\zcat)/(1+\zcat)$). The orange circles and the solid line show the median distribution. The error bars indicate $1\sigma$ scatter. (Right) Similar to the left panels, but comparing \texttt{RVSNUpy} redshifts measured from Hectospec spectra based on the Hectospec templates ($z\mathrm{^{RVSNUpy}_{Hecto, Hecto}}$) and $z\mathrm{^{SDSScat}}$ ($c\Delta z /(1+z) = c(\zpyhectohecto-\zcat)/(1+\zcat)$).}
    \label{fig3}
\end{figure}

Figure \ref{fig3} compares the SDSS redshifts ($\zcat$) with the \texttt{RVSNUpy} redshifts we derived from the SDSS spectra with SDSS templates and Hectospec spectra with Hectospec templates (i.e., $\zpysdsssdss$ and $\zpyhectohecto$, respectively). Figure \ref{fig3} (a) shows that $\zpysdsssdss$s are generally consistent with $\zcat$ over the entire redshift range ($z < 0.8$) with a small systematic offset. The typical redshift difference is $10.9 \pm 12.6~\kms$. Figure \ref{fig3} (b) shows that $z^{\rm RVSNUpy}_{\rm Hecto, Hecto}$s are systematically offset from $z^{\rm SDSS~cat}$ with a typical offset of  $-37.3 \pm 30.7~\kms$. The offset increases with redshift for $z < 0.25$ and remains constant at $z > 0.25$. The redshift offset and trend for the Hectospec spectra are consistent with the results reported by \citet{Sohn21} (and also \citealp{Geller14, Damjanov18}).  

\subsection{The origin of SDSS and Hectospec redshift offsets}
\label{sec:offsetorigin}
Here, we investigate the origin of the systematic offset between the SDSS redshifts from the SDSS catalog (i.e., $\zcat$) and the Hectospec redshifts measured with \texttt{RVSNUpy} based on Hectospec templates (i.e., $\zpyhectohecto$). We emphasize that the two redshifts we test here are not determined based on the same tool. The offset between these two redshifts could be explained by one or more of the following; The first possibility is that the offset corresponds with the offset between the observed SDSS and Hectospec spectra for the same objects. Another possible explanation is the use of different redshift measurement tools for the redshift estimation. Finally, the offset could be caused by differences in the redshift measurement template spectra used for the SDSS and Hectospec observations.

\begin{figure}
    \centering
    \includegraphics[width=.5\textwidth]{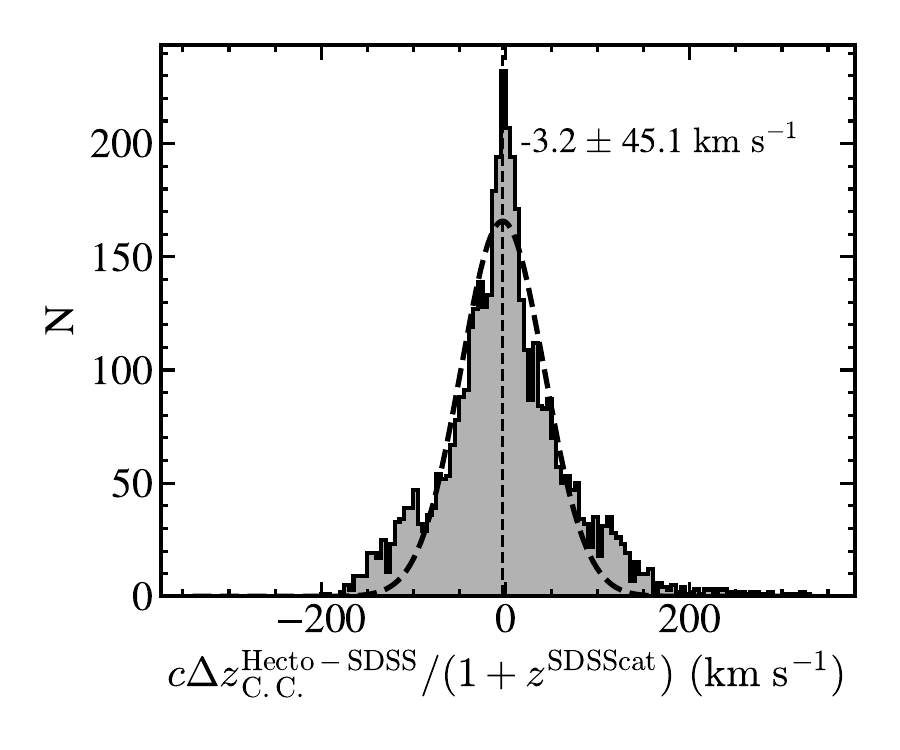}
    \caption{The distribution of radial velocity offsets derived from performing cross-correlation between the SDSS and Hectospec spectra. The black dashed line shows the best-fit Gaussian with a mean of $-3.2~\kms$}
    \label{fig4}
\end{figure}

Figure \ref{fig4} shows the radial velocity offset ($c\Delta z\mathrm{^{Hecto-SDSS}_{C.C.}}/(1+\zcat)$) between the SDSS and Hectospec spectra for the HectoMAP test objects measured through the cross-correlation between them. The mean of the radial velocity offsets is $c\Delta z\mathrm{^{Hecto-SDSS}_{C.C.}}/(1+\zcat) = -3.2 \pm 45.1~\kms$, indicating that the offset between SDSS and Hectospec spectra for the same objects is negligible. Thus, the redshift offset does not originate from the difference in SDSS and Hectospec observed spectra. 

\begin{figure}[h]
    \centering
    \includegraphics[width=.7\linewidth]{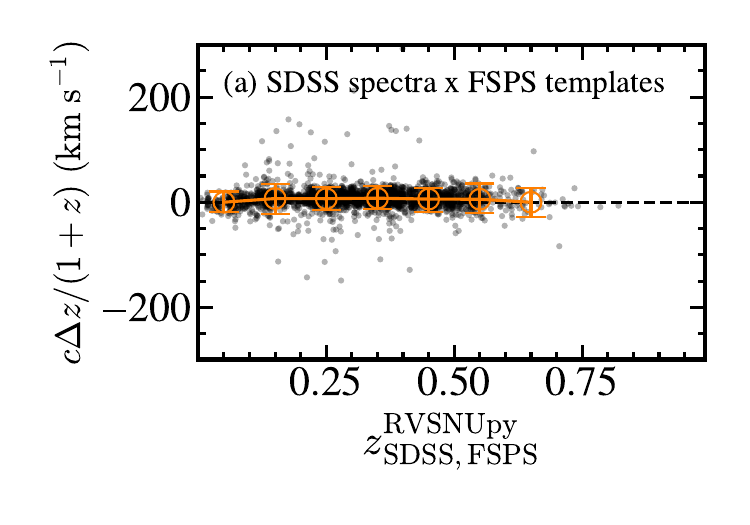}
    \caption{(Left) Comparison between RVSNUpy redshifts measured from SDSS spectra based on the \texttt{FSPS} templates (i.e., $z\mathrm{^{RVSNUpy}_{SDSS, FSPS}}$) and $z\mathrm{^{SDSScat}}$ for the HectoMAP test sample ($c\Delta z /(1+z) = c(\zpysdssfsps-\zcat)/(1+\zcat)$). The orange circles and the solid line show the median distribution. The error bars indicate $1\sigma$ scatter.}
    \label{fig5}
\end{figure}

Another potential cause of the redshift offset is the use of different redshift measurement tools. We measure redshifts using \texttt{RVSNUpy} with the same combination of observed spectra and template spectra as the $\zcat$ measurement based on the full spectral fitting (i.e. SDSS spectra and \texttt{FSPS} templates ($\zpysdssfsps)$). Figure \ref{fig5} shows that $\zpysdssfsps$ and $\zcat$ are relatively consistent with a very small offset ($c\Delta z = 6.8~\kms$). Therefore, the choice of redshift measurement tool is not the source of the redshift offset. 

\begin{figure}[h]
    \centering
    \includegraphics[width=.8\linewidth]{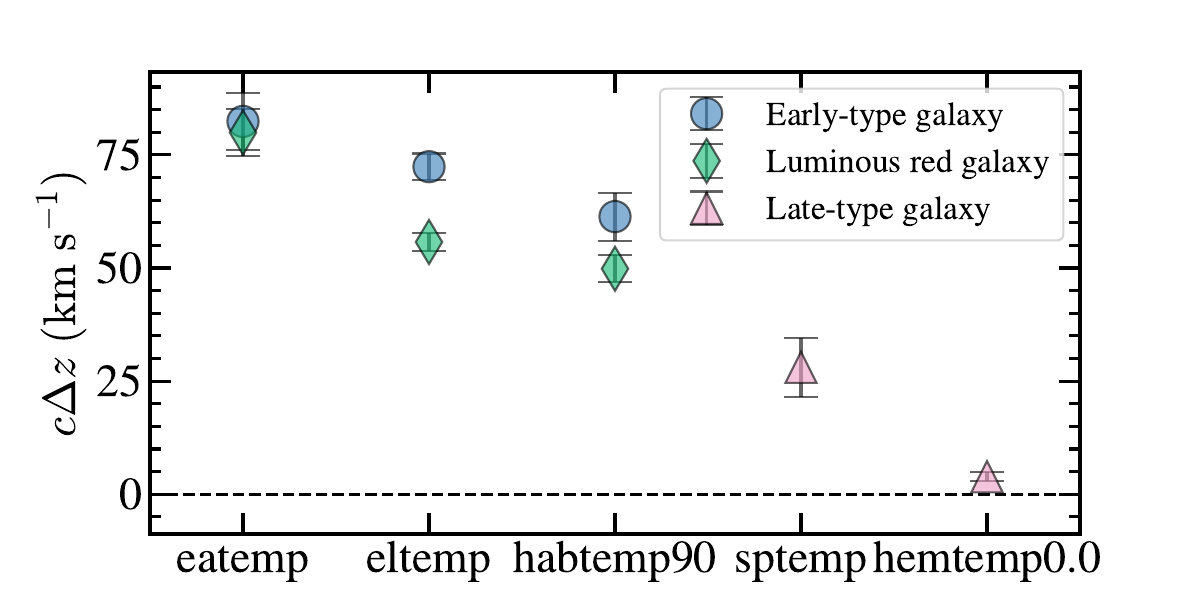}
    \caption{The radial velocity offsets of Hectospec templates from the SDSS templates. The horizontal axis indicates the various Hectospec templates. The sky blue circles, green diamonds, and pink triangles display the radial velocity offset of the Hectospec templates from the `Early-type galaxy', `Luminous red galaxy', and `Late-type galaxy' of the SDSS templates, respectively. $c\Delta>0$ indicates the Hectospec templates are redshifted from the SDSS templates (i.e., $c\Delta z=c(z_{\rm Hectospec~templates}-z_{\rm SDSS~templates})$).}
    \label{fig6}
\end{figure}

\begin{figure}[h]
    \centering
    \includegraphics[width=.6\linewidth]{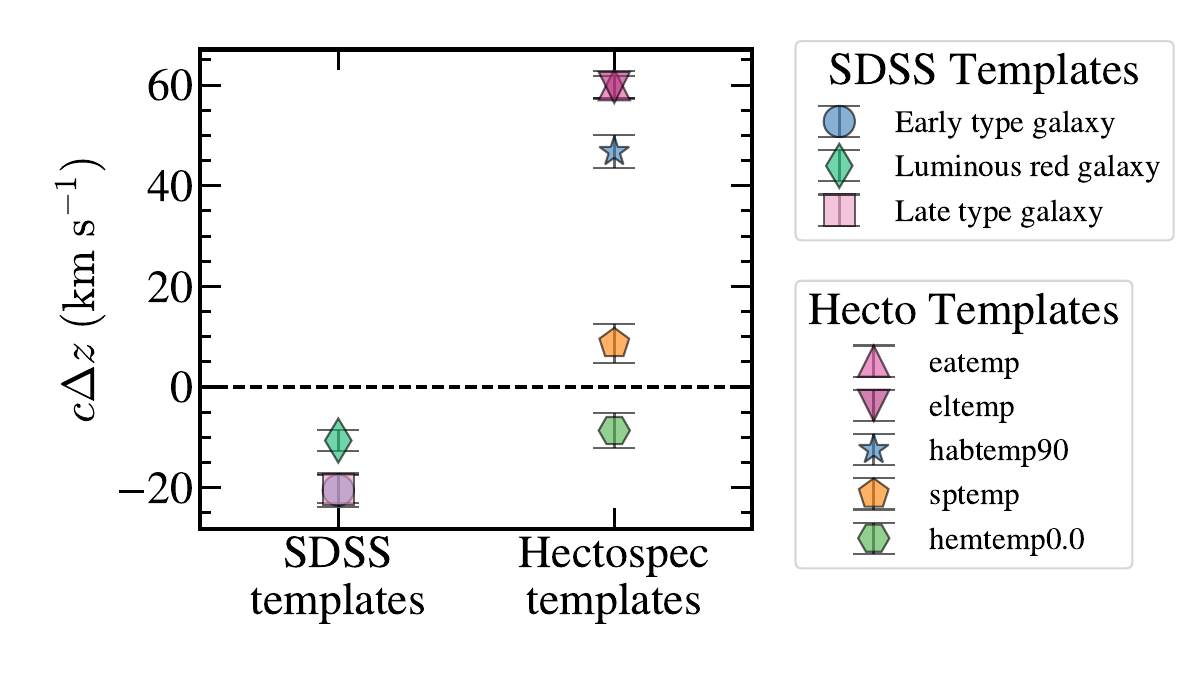}
    \caption{The radial velocity offsets of the SDSS and Hectospec templates with respect to the \texttt{FSPS} templates in the rest frame. $c\Delta z/(1+z)>0$ indicates the SDSS/Hectospec templates are redshifted from the \texttt{FSPS} templates (i.e., $c\Delta z=c(z_{\rm SDSS/Hectospec~templates}-z_{\rm FSPS~templates})$).}
    \label{fig7}
\end{figure}

\begin{figure}[h]
    \centering
    \includegraphics[width=\linewidth]{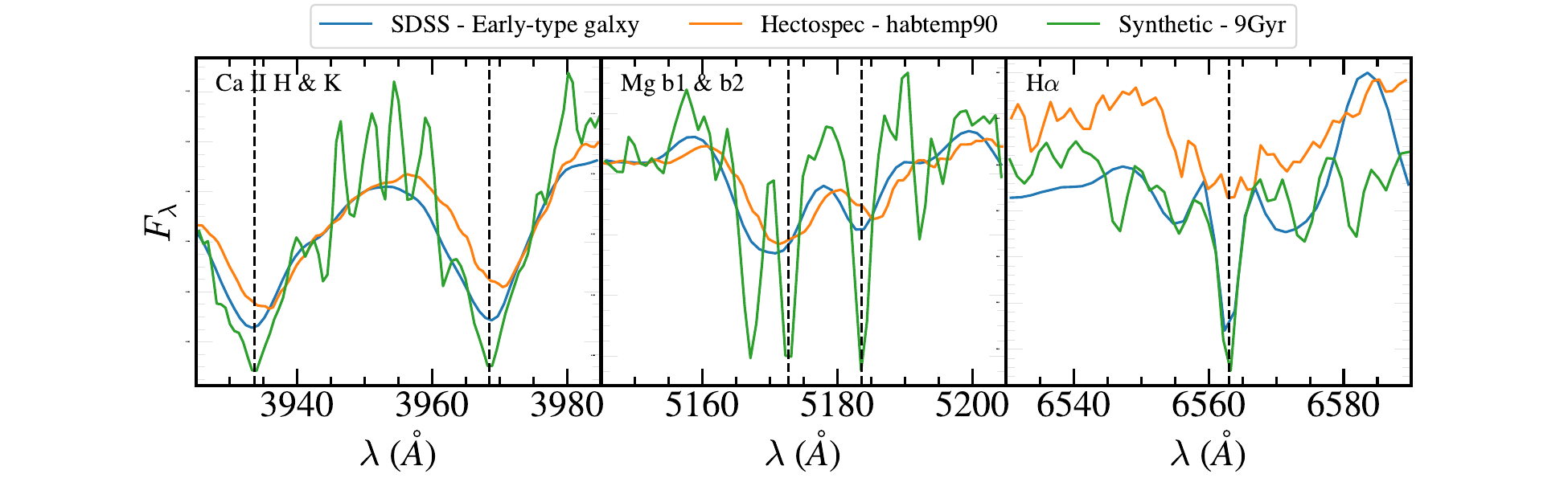}
    \caption{The various template spectra around distinctive spectral lines. From the left to right, each panel shows the template spectra around Ca H\&K, Mg b1\&b2, and H${\rm \alpha}$ lines. The skyblue, orange, and green lines are the `Earl-type galaxy' of the SSDS templates, `habtemp90' of the Hectospec templates, and `9 Gyr' SSP spectrum of the \texttt{FSPS} templates, respectively. The vertical dashed lines indicate the wavelength of the spectral lines in the rest frame.}
    \label{fig8}
\end{figure}

\begin{figure}[h]
    \centering
    \includegraphics[width=.7\linewidth]{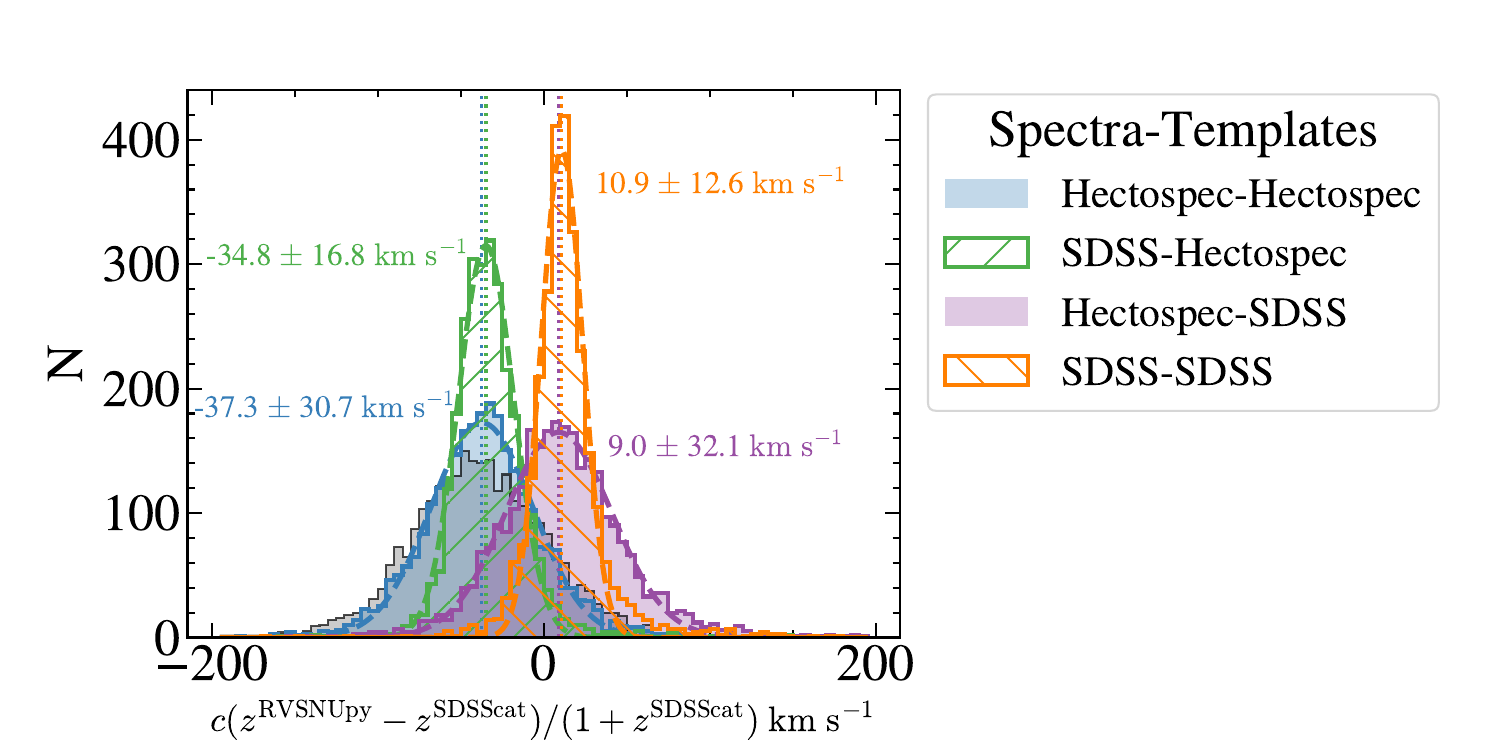}
    \caption{Distributions showing the difference between \texttt{RVSNUpy} redshifts measured from various observed spectra/template spectra combinations and the redshifts from the SDSS catalog ($z^{\rm SDSScat}$). The blue and green histograms show the redshift measurements based on Hectospec and SDSS spectra cross-correlated with the Hectospec templates. The magenta and orange histograms show the measurements based on Hectospec and SDSS spectra cross-correlated with the SDSS templates. The gray histogram behind the blue histogram indicates the redshift offset between Hectospec and SDSS spectra reported by \citet{Sohn21}. Dashed lines with different colors show the best-fit Gaussian for the distributions with the corresponding colors. Vertical lines indicate the peak locations (marked with the numbers).}
    \label{fig9}
\end{figure}

Next, we investigate the impact of the template selection on the redshift measurements. To check if the SDSS and Hectospec templates are consistent with one other, we cross-correlate these template spectra as we did for the observed spectra. Figure \ref{fig6} shows the radial velocity differences between the SDSS and Hectospec templates. Interestingly, the SDSS and Hectospec templates for similar galaxy types show offsets in their redshift velocities. For example, the `eatemp' Hectospec template representing the E+A type galaxy spectrum shows a $c\Delta z \sim 82~\kms$ offset from the `Early-type galaxy' SDSS template in the cross-correlation signal. The offsets between the redshift templates presumably propagate to the redshift measurements, resulting in the redshift offset. 

We then cross-correlate both the SDSS and Hectospec templates with the \texttt{FSPS} templates which are theoretical synthetic spectra in the rest-frame. Figure \ref{fig7} displays the offset between SDSS and Hectospec templates relative to the \texttt{FSPS} templates. In general, the SDSS templates show small offsets ($\sim10-20~\kms$) compared to the \texttt{FSPS} templates. However, the Hectospec templates are significantly redshifted compared to the \texttt{FSPS} templates. In particular, the offset is much larger ($>40~\kms$) for the absorption line templates (`eatemp', `eltemp', `habetmp90') and is much smaller ($<20~\kms$) for the emission line templates (`sptemp' and `hemtemp0.0'). 

Figure \ref{fig8} displays `Early-type galaxy' of the SDSS templates (skyblue), `habtemp90' of the Hectospec teampltes (orange), and `9 Gyr SSP spectrum' of the \texttt{FSPS} templates (green) around the spectral lines. The dashed lines mark the rest-frame wavelengths of the spectral lines. The spectral lines of `Early-type galaxy', `habtemp90', and `9 Gyr SSP spectrum' shows different shapes and center wavelengths. In particular, the spectral liens of `habtmp90' significantly deviates from the rest frame line wavelengths, indicating that the offset of the  Hectospec templates from the rest frame. These velocity offsets result in the measured redshift offsets from Hectospec spectra in Figure \ref{fig3}. 

Figure \ref{fig9} demonstrates the impact of the template spectra. We plot the differences between $\zcat$ and \texttt{RVSNUpy} redshifts for the HectoMAP test sample based on four combinations of the observed spectra and template spectra: Hectospec $\times$ Hectospec templates, Hectospec $\times$ the SDSS templates, SDSS $\times$ Hectospec templates, and SDSS $\times$ the SDSS templates. Because we use \texttt{RVSNUpy} for all combinations, the difference in the redshift measurement techniques is irrelevant to the result here.

The orange histogram in Figure \ref{fig9} displays the offset between $\zcat$ and $\zpysdsssdss$ and shows a systematic offset of $10.9~\kms$. The purple histogram ($c(\zpyhectosdss - \zcat)/(1+\zcat)$) shows a systematic offset of $9.0~\kms$, similar to the orange histogram ($c(\zpysdsssdss - \zcat)/(1+\zcat)$). In contrast, both the green and blue histograms, which show the redshift measurements based on the Hectospec templates, are offset by $\sim 35~\kms$ from $\zcat$, similar to the redshift offset reported by \citet{Sohn21} (gray). This is a clear indicator that the offset in template spectra is the primary source of the redshift offset. Thus, the choice of template spectra is important for measuring homogeneous redshift measurements to compare and compile data from other massive spectroscopic surveys. 

\subsection{RVSNUpy with the universal templates} \label{sec:obsconclusion}

Cross-correlation between templates and synthesized spectra reveals slight offsets of $10{-}20~\kms$ for SDSS templates and $10{-}70~\kms$ for Hectospec templates relative to the rest frame. In particular, the offsets of the Hectospec absorption line templates from the rest frame are significant and thus result in systematic offsets in the redshift measurements. To reduce the systematic redshift offset introduced by the offsets in the template spectra, we require a universal set of template spectra that enable homogeneous redshift measurements. 

We constructed a new set of template spectra based on the SDSS templates. We decide to use the empirical SDSS templates shifted to the rest-frame rather than using the synthetic \texttt{FSPS} templates. Because the synthetic SSP spectra do not reflect the variation in the underlying stellar populations and the observational effect, the cross-correlation signals for the observed spectra and synthetic SSP templates are generally weaker than those of empirical template spectra. To calibrate the radial velocity offset of the SDSS templates shown in Figure \ref{fig7} and \ref{fig8}, we shifted three SDSS templates (i.e., `Early Type Galaxy', `Luminous Red Galaxy', and `Late Type Galaxy') along the wavelength direction to match the \texttt{FSPS} templates. Hereafter, we refer to these three template spectra as the `universal templates' and the redshifts we derive with these template spectra are denoted as $z\mathrm{^{RNSNUpy}_{SDSS/Hecto, utemp}}$. 

\begin{figure}[h]
    \centering
    \includegraphics[width=\linewidth]{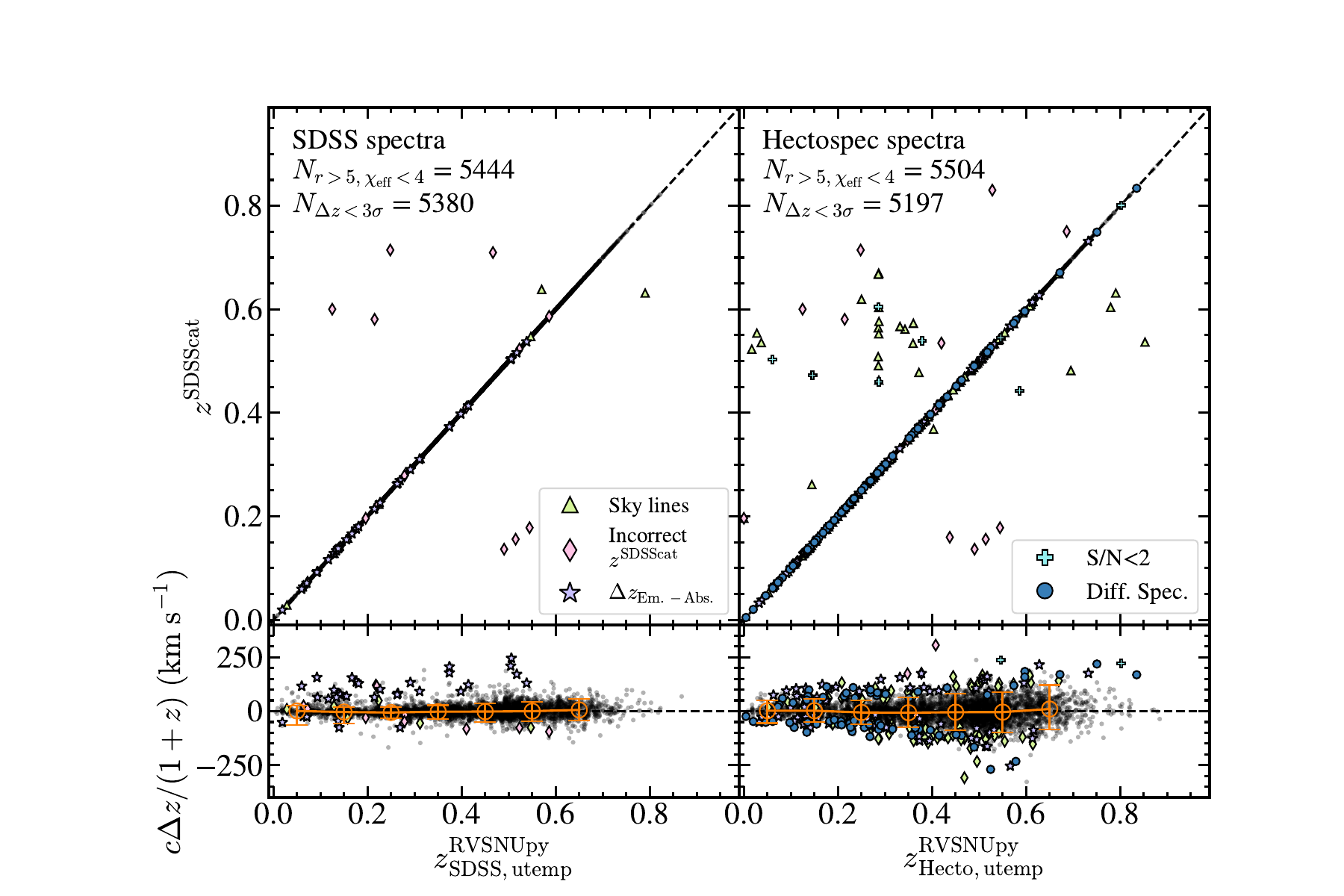}
    \caption{(Left) Comparison between RVSNUpy redshifts measured from SDSS spectra (i.e., $z\mathrm{^{RVSNUpy}_{SDSS, utemp}}$)  and SDSS redshifts ($z\mathrm{^{SDSScat}}$) for the HectoMAP test sample. The upper panel shows a one-to-one comparison, and the lower panel shows the redshift difference ($c\Delta z/(1+z)=c(\zpysdssutemp-\zcat)/(1+\zcat)$) as a function of $z\mathrm{^{RVSNUpy}_{SDSS, utemp}}$. Gray circles display individual redshift measurements. Light green triangles, pink diamonds, and purple stars mark outliers with a redshift difference ($\Delta z$) larger than three times the uncertainty of the redshift difference due to poorly subtracted sky lines, incorrect $\zcat$,  and the redshift difference between absorption and emission lines, respectively. In the lower panel, orange circles and the solid line show the median distribution. The error bars indicate $1\sigma$ scatter. (Right) Similar to the left panels, but comparing \texttt{RVSNUpy} redshifts measured from Hectospec spectra ($z\mathrm{^{RVSNUpy}_{Hecto, utemp}}$) and $z\mathrm{^{SDSScat}}$. Cyan plus signs and blue circles mark outliers due to low S/N and intrinsic differences between Hectospec and SDSS spectra.}
    \label{fig10}
\end{figure}

Figure \ref{fig10} compares the SDSS redshifts ($\zcat$) with RVSNUpy redshifts derived using universal templates. Among the 6009 HectoMAP comparison sample, we measure $r > 5,~\chi^2_{\rm eff}<4$ redshifts from 5444 SDSS spectra and 5504 Hectospec spectra. The \texttt{RVSNUpy} redshifts measured with the universal templates are essentially identical to the corresponding redshifts in $\zcat$ with a very small fraction of outliers. The lower panels display the redshift difference as a function of $\zpysdssutemp$. Orange circles and their error bars in the lower panels show the median and 1$\sigma$ scatter of redshift difference with respect to the SDSS redshifts: $-2.8\pm17.3~\kms$ for $z^{\rm RVSNUpy}_{\rm SDSS, utemp}$ and $-3.1\pm38.6~\kms$ for $z^{\rm RVSNUpy}_{\rm Hecto, utemp}$ at $z < 0.8$. These offsets are smaller than the redshift offsets based on the measurements of \texttt{RVSAO} for the HectoMAP sample (\citealp{Sohn23}; $-41.8\pm45.7~\kms$).

In Figure \ref{fig10}, various symbols mark the outliers with a significant redshift difference:  $\Delta z > 3 \delta({\Delta z})$, where $\delta \Delta z$ indicates the uncertainty of the redshift difference (i.e. $\sqrt{(\delta z^{\rm RVSNUpy}_{\rm SDSS/Hecto, utemp})^2+(\delta\zcat)^2}$). There are 64 ($\sim 1.1\%$) outliers in the SDSS spectra and 307 ($\sim 5.6\%$) outliers in the Hectospec spectra from the HectoMAP test sample. 

We visually inspected the SDSS and Hectospec spectra for outliers. Some outliers arise from poorly sky-subtracted spectra (light green triangles) or low S/N ratios (cyan plus signs). Due to the low quality of the spectra, \texttt{RVSNUpy} cannot find the peaks in the cross-correlation signals. In particular, most outliers with extreme redshift differences are a result of low-quality spectra.  

\begin{figure}
    \centering
    \includegraphics[width=0.9\linewidth]{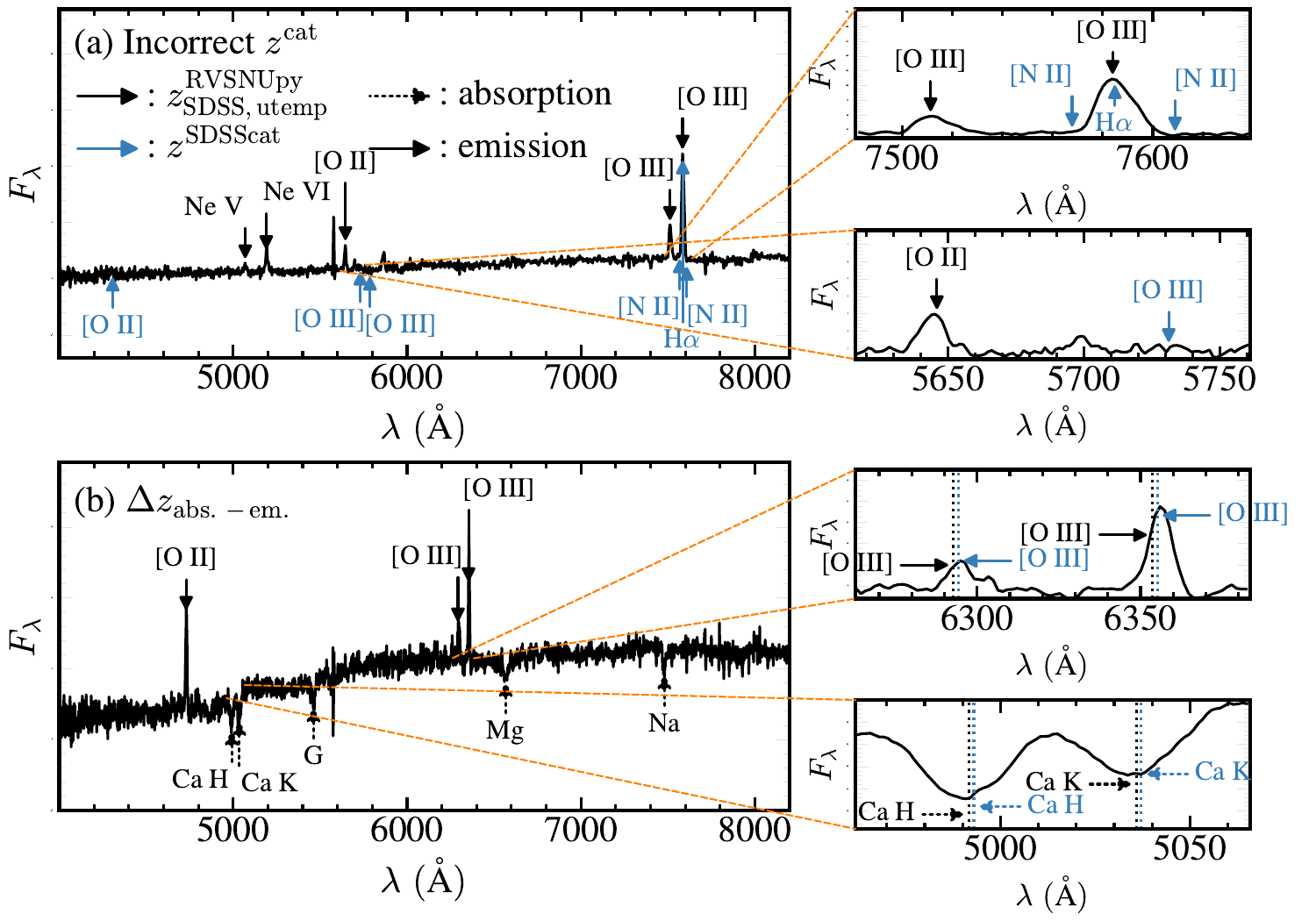}
    \caption{Example spectra of (a) incorrect $\zcat$ and (b) difference in redshifts between absorption and emission lines. The black and sky blue arrows with text display the locations of spectral lines from $z^{\rm RVSNUpy}$ and $\zcat$. The solid and dotted lines indicate absorption and emission lines, respectively. The small panels on the right side show the spectra in the small wavelength range where strong lines exist.}
    \label{fig11}
\end{figure}

For $\sim$23\% of SDSS spectra and $\sim$6\% of Hectospec spectra outliers, \texttt{RVSNUpy} redshifts appear more reasonable than the redshift listed in the SDSS catalog. Figure \ref{fig11} displays (a) an example Hectospec spectrum of an outlier ($c\Delta z / (1+z) = 71081~\kms$). Black and sky blue arrows indicate spectral lines shifted by $z^{\rm RVSNUpy}_{\rm Hecto, utemp}$ and $\zcat$, respectively. In the example spectrum, there is no hint of [O III] emission at the expected wavelength based on $\zcat$. Furthermore, the [N II]$\lambda 6548$ and [N II]$\lambda 6583$ emission lines do not appear at the expected wavelength, assuming $\zcat$ is the true redshift. In this case, the \texttt{RVSNUpy} redshift is consistent with the locations of multiple emission lines. 

There are also outliers that display both absorption and emission lines in their spectra ($\sim$50\% for the SDSS spectra and 24\% for the Hectospec spectra). Figure \ref{fig11} shows (b) an example SDSS spectrum of an outlier (c$\Delta z / (1+z) = -75.3~\kms$). The \texttt{RVSNUpy} redshifts measured based on absorption and emission line templates differ by $-77.94~\kms$. The $z^{\rm RVSNUpy}_{\rm SDSS/Hecto, utemp}$ based on the emission line template is consistent with $\zcat$. However, we designed \texttt{RVSNUpy} to choose the redshift measured based on the absorption line template with a higher priority. We choose the redshift measured based on absorption lines because the absorption lines presumably trace the motion of the stellar components of the galaxies well. The zoomed-in sections of Figure \ref{fig11} (b) show the Ca H and K lines demonstrate that the \texttt{RVSNUpy} redshift determines the location of the absorption line successfully. Therefore, we trust the \texttt{RVSNUpy} redshifts determined based on absorption lines for outliers with absorption and emission lines.

\begin{figure}[h]
    \centering
    \includegraphics[width=.8\linewidth]{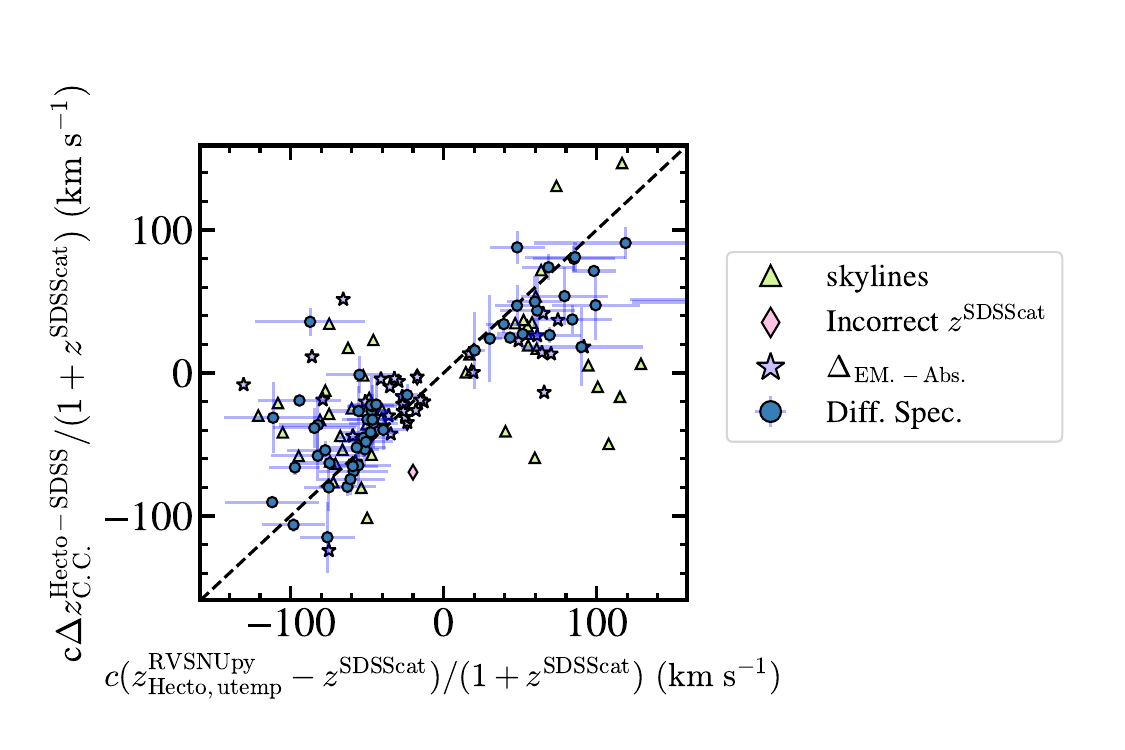}
    \caption{The radial velocity differences estimated from the cross-correlation between SDSS and Hectospec spectra as a function of the normalized difference between $\zpyhectoutemp$ and $\zcat$. Blue circles show the cases in which SDSS and Hectospec spectra are intrinsically different. Light green triangles, pink diamonds, and purple stars display outliers where skylines are poorly subtracted, $\zcat$ is incorrect, and the redshifts of absorption and emission lines are different.}
    \label{fig12}
\end{figure}

Finally, $\sim$58\% of the outliers in the Hectospec spectra display a relative radial velocity offset smaller than $250~\kms$ which results from the intrinsic deviations between the SDSS and Hectospec spectra. The spectra of these outliers have a high S/N (i.e., $> 3$) and lack strong emission lines to hinder \texttt{RVSNUpy}. Figure \ref{fig12} displays the radial velocity difference derived from the cross-correlation between SDSS and Hectosepc spectra (i.e., $c\Delta z\mathrm{^{Hecto-SDSS}_{C.C.}}/(1+\zcat)$) as a function of the difference between $\zpyhectoutemp$ and $\zcat$. Most of the outliers (shown as blue circles) indeed show a large radial velocity difference in the observed spectra, highlighting the intrinsic differences between the observed spectra. Thus, the redshift difference results from the intrinsic differences in the observed spectra. 

We show that \texttt{RVSNUpy} successfully measures the redshifts of both SDSS and Hectospec spectra with the universal templates. \texttt{RVSNUpy} redshifts are generally consistent with the redshifts from the SDSS catalog. For the outliers, we carefully show that 1) \texttt{RVSNUpy} redshifts explain the observed spectra better or 2) the SDSS and Hectospec spectra are intrinsically shifted by the redshift offset. Based on this test, we affirm that \texttt{RVSNUpy} yields redshift measurements without any significant systematic bias. Thus, \texttt{RVSNUpy} can become an efficient and powerful redshift measurement tool for future large-scale spectroscopic surveys. 

\section{Conclusion} \label{sec:conclusion}

In this paper, we develop \texttt{RVSNUpy}, a new spectroscopic redshift measurement Python package. \texttt{RVSNUpy} measures spectroscopic redshifts based on the cross-correlation technique between observed spectra and redshift template spectra. The cross-correlation technique is a simple but powerful tool that can efficiently measure redshifts of large-scale data sets. \texttt{RVSNupy} is a universal Python package based on inverse variance weighted cross-correlation. 

We described how \texttt{RVSNUpy} derives the redshifts based on the inverse variance weighted cross-correlation method. The processes of \texttt{RVSNUpy} include preprocessing the input spectrum, performing a cross-correlation between the input spectrum normalized by the continuum and the redshift template spectra, and determining redshift from the cross-correlation signal. Through iteration of this process over various types of the template spectra, both absorption and emission-line dominated, \texttt{RVSNUpy} derives the final redshift of the input spectra based on the significance of the cross-correlation signal. 

We tested the performance of \texttt{RVSNUpy} in three steps. First, we tested the redshifts of the synthetic spectra with a simple stellar population model. \texttt{RVSNUpy} reproduces the redshifts successfully regardless of the age, spectral resolution, or signal-to-noise ratio used for creating the synthetic spectra. 

We investigated the impact of template choice in the redshift measurements. For this test, we used the HectoMAP test sample, including $\sim6000$ objects of both SDSS and MMT/Hectospec spectra. The redshifts we derived from SDSS and Hectospec spectra are generally consistent but show a slight systematic offset ($\sim 10{\rm~or~}35~\kms$). This systematic offset does not result from the intrinsic difference between SDSS and Hectospec spectra. We identified that the redshift offsets are resolved when we use a homogeneous set of template spectra regardless of the observed spectra we use. In other words, the redshifts we derived based on the cross-correlation between SDSS and Hectospec observed spectra with the SDSS (or Hectospec) templates agree without any systematic offset. This test result strongly suggests that the systematic redshift offset originates from the differences between the SDSS and Hectospec templates. Indeed, the Hectospec templates are slightly offset from rest frame, which results in the redshift offset. 

Finally, we tested the performance of \texttt{RVSNUpy} based on the universal templates we constructed to reproduce the widely used SDSS spectroscopic redshifts. The universal templates include SDSS templates shifted to rest-frame based on the redshift offset we derived from the cross-correlation between the original SDSS templates and synthetic spectra. The final redshifts of the HectoMAP test sample we derived using the universal templates agree with the redshifts from the SDSS catalog.

\texttt{RVSNUpy} is a useful package that can derive redshifts from extensive datasets such as upcoming large-scale spectroscopic surveys (e.g., A-SPEC, DESI, 4MOST, Subaru/PFS). The cross-correlation technique is more efficient than the full spectral fitting technique, and it measures empirical redshifts without suffering from possible systematic biases introduced by the model spectra generated for the full spectral fitting. In conclusion, we wish to highlight that \texttt{RVSNUpy} is a simple, powerful, and straightforward tool for deriving accurate spectroscopic redshifts from large surveys. 

\section*{Acknowledgements}
We appreciate the referee for the insightful and constructive review of our manuscript. We thank Margaret Geller and Scott Kanyon for their insightful suggestions on employing cross-correlation in real space, as well as their discussions and comments that greatly improved this manuscript. This work was supported by the National Research Foundation of Korea (NRF) grant funded by the Korea government (MSIT) (RS-2023-00210597). This work was supported by Creative-Pioneering Researchers Program through Seoul National University. JS is also supported by a Global-LAMP Program of the NRF grant funded by the Ministry of Education (No. RS-2023-00301976). HSH acknowledges the support of the National Research Foundation of Korea (NRF) grant funded by the Korea government (MSIT), NRF-2021R1A2C1094577, Samsung Electronic Co., Ltd. (Project Number IO220811-01945-01), and Hyunsong Educational \& Cultural Foundation. 

\bibliography{ms}{}
\bibliographystyle{aasjournal}

\end{document}